\documentclass[openacc]{rsproca_new}
\usepackage[matrix,frame,arrow]{xy}
\usepackage[matrix,frame,arrow]{xy}
\usepackage{amsmath}
\newcommand{\qw}[1][-1]{\ar @{-} [0,#1]}

\newcommand{\gate}[1]{*{\xy *+<.6em>{#1};p\save+LU;+RU **\dir{-}\restore\save+RU;+RD **\dir{-}\restore\save+RD;+LD **\dir{-}\restore\POS+LD;+LU **\dir{-}\endxy} \qw}

\newcommand{\multigate}[2]{*+<1em,.9em>{\hphantom{#2}} \qw \POS[0,0].[#1,0];p !C *{#2},p \save+LU;+RU **\dir{-}\restore\save+RU;+RD **\dir{-}\restore\save+RD;+LD **\dir{-}\restore\save+LD;+LU **\dir{-}\restore}

\newcommand{\ghost}[1]{*+<1em,.9em>{\hphantom{#1}} \qw}

\newcommand{\ustick}[1]{*!D!<0em,-.5em>=<0em>{#1}}

\newcommand{\Qcircuit}[1][0em]{\xymatrix @*=<#1>}

\newcommand{\pureghost}[1]{*+<1em,.9em>{\hphantom{#1}}}

\usepackage{cleveref}
\usepackage{booktabs}

\usepackage{epsfig,amsmath,amssymb,mathtools,amscd}
\usepackage[active]{srcltx}
\usepackage{dsfont}
\usepackage{color}

\newtheorem{lemma}{Lemma} 

\newtheorem{proposition}{Proposition}

\newtheorem{corollary}{Corollary} 
\newtheorem{theorem}{Theorem}
 
\newtheorem{definition}{Definition}

\def\Proof{\medskip\par\noindent{\bf Proof. }}
\def\choi{\mathsf{Ch}}

\def\qed{$\,\blacksquare$\par}

\newcommand{\spn}{\operatorname{span}}
\def\Tr{\operatorname{Tr}}
\def\dim{\operatorname{dim}}

\def\bvec#1{\mathbf{#1}}
\def\>{\rangle}
\def\<{\langle}

\def\Ch{\mathsf{Ch}}

\def\T+{\mathsf{T}_+}
\def\ETy{\set{EleTypes}}
\def\Ty{\set{Types}}

\def\Ev#1{\set{T}_{\mathbb R}(#1)}
\def\Evd#1{\set{T}_1(#1)}
\def\Eva#1{\set{T}(#1)}
\def\Evp#1{\set{T}_+(#1)}
\def\trasf#1{\mathcal{#1}}
\def\tI{\trasf I}

\def\para{\parallel}

\newcommand\Item[1][]{%
  \ifx\relax#1\relax  \item \else \item[#1] \fi
  \abovedisplayskip=0pt\abovedisplayshortskip=0pt~\vspace*{-\baselineskip}}

\newcommand{\ket}[1]{|#1\rangle}

\newcommand{\Bra}[1]{\langle \! \langle#1|}
\newcommand{\Ket}[1]{|#1\rangle \! \rangle}

\newcommand{\KetBra}[2]{{\Ket{#1}\Bra{#2}}}
\newcommand{\hilb}[1]{\mathcal{#1}}
\newcommand{\set}[1]{{\sf #1}}

\newcommand{\Lin}[1]{\mathcal{L}(#1)}

\def\transf#1{\scriptstyle{#1}}
\def\tC{\transf{C}}

\begin{document}

\title{Theoretical framework for Higher-Order Quantum Theory}

\author{
Alessandro Bisio$^{1}$ and Paolo Perinotti$^{2}$}

\address{$^{1}$QUIT group, Dipartimento di
  Fisica, Universit\`{a} degli Studi di Pavia, and INFN, Gruppo IV , via Bassi 6, 27100 Pavia, Italy\\
$^{2}$QUIT group, Dipartimento di
  Fisica, Universit\`{a} degli Studi di Pavia, and INFN, Gruppo IV , via Bassi 6, 27100 Pavia, Italy}

\subject{Quantum foundations, Quantum computation, Quantum information, }


\corres{Paolo Perinotti\\
\email{paolo.perinotti@unipv.it}}

\begin{abstract}
  Higher-order quantum  {theory} is an extension of quantum
  theory where one introduces transformations whose input and
  output are transformations, thus generalizing the notion of channels
  and quantum operations. The generalization then goes recursively,
  with the construction of a full hierarchy of maps of increasingly
  higher order. The analysis of special cases already showed that
  higher-order quantum functions exhibit features that cannot be tracked down
  to the usual circuits, such as indefinite causal structures,
  providing provable advantages over circuital maps. The present
  treatment provides a general framework where this kind of analysis
  can be carried out in full generality.  {The hierarchy of higher-order quantum maps} is introduced axiomatically with a formulation based on
  the language of types of transformations. Complete positivity of
  higher-order maps is derived from the general admissibility
  conditions instead of being postulated as in previous
  approaches. The recursive characterization of convex sets of maps of
  a given type is used to prove equivalence relations between
  different types. The axioms  {of the framework} do not
  refer to the specific mathematical structure of quantum theory, and
  can therefore be exported in the context of any operational
  probabilistic theory.
\end{abstract}


\maketitle
  \section{Introduction}

  The key idea behind the  {higher-order quantum theory}
  is the promotion of quantum channels, which are normally
  considered as the logical gates in a quantum circuit, to
  the role of inputs, thus introducing ``second-order" gates
  that transform channels into channels. In order to analyse
  this kind of question, the first step to take is to
  consider the channels as exquisitely mathematical objects,
  and then define the class of maps from the set of channels
  to itself. This class must satisfy some minimal {\em
    admissibility} requirement, that are the loosest
  constraints for the maps to respect the probabilistic
  structure of quantum theory, as in the axiomatisation of
  completely positive maps \cite{Kraus}. An admissible map
  from quantum operations to quantum operations must then i)
  respect convex combinations (i.e.~it must be linear), ii)
  respect the set of channels also when applied locally to
  bipartite channels, and iii) preserve the normalization of
  channels. In Ref.~\cite{chiribella2008transforming} it was
  proved that this kind of transformations precisely
  corresponds to inserting the input channel into a fixed
  open circuit as in the following diagram

\begin{align}
  \begin{aligned}
\Qcircuit@C=0.5em @R=1em {
    &\ustick{\scriptstyle{A}}&\gate{{R}}&\ustick{\scriptstyle{B}}\qw&}\\
  \end{aligned}\ \mapsto\ 
  \begin{aligned}
\Qcircuit@C=0.5em @R=1em {
   &\ustick{\scriptstyle{C}}\qw&\multigate{1}{{\phantom{\tC}}}& \ustick{\scriptstyle{A}}\qw&\gate{R}&\ustick{\scriptstyle{B}}\qw&\multigate{1}{\phantom{\tC}}&\ustick{\scriptstyle{D}}\qw\\
   &&\pureghost{\tC}  &\qw&\qw&\qw&\ghost{\tC}&
}
  \end{aligned}\ .
\label{eq:supm}
\end{align}

This idea is then immediately brought to its most general
scenario: every kind of map can be raised to the level of
the input of a computation at a further level in the
hierarchy. Such a construction is not exclusive to quantum
computation, and can be made also in the case of classical
gates
\cite{1402-4896-2014-T163-014013,baumeler2014perfect}. Actually,
the first instance of higher-order computation can be
tracked back to the invention of Lambda calculus. A quantum
version constituting a model for higher-order quantum
computation was elaborated in
Ref.~\cite{selinger_valiron_2009}.

A relevant sub-hierarchy of maps is the one consisting of
{\em quantum combs}, that can be thought of as the
generalization of maps of eq.~\eqref{eq:supm} with more than
two ``teeth'', where one comb with $n$ teeth maps a comb
with $n-1$ teeth to a channel. This hierarchy was
extensively studied in the last decade (for exhaustive
reviews see
\cite{PhysRevLett.101.060401,PhysRevA.80.022339,bisio2016quantum,Gutoski:2007:TGT:1250790.1250873}). The
distinctive feature of maps in this sub-hierarchy is that
they can be implemented by modular connection of networks of
quantum gates.

As soon as one makes one step further, e.g.~considering
transformations from combs to combs, maps that cannot
be implemented by a quantum circuit appear
\cite{PhysRevA.80.022339,PhysRevA.88.022318}.   {A
  paradigmatic example is the quantum SWITCH map
  \cite{PhysRevA.88.022318} which takes as an input two
  quantum channels, say $\mathcal{A} $ and
  $\mathcal{B}$, and outputs the coherent superposition
  of the sequential applications of the two channels in
  two different order, i.e.
  $\mathcal{A} \circ \mathcal{B}$ and
  $\mathcal{B} \circ \mathcal{A}$. } In some special
case, these maps can be thought of as mixtures or
``superpositions" of causally ordered circuits
\cite{PhysRevA.88.022318,Oreshkov:2012aa}, as
precognized in the pioneering proposals of Hardy
\cite{Hardy:2007aa}. Important results followed in the
subsequent years, showing advantages over standard
quantum computation in non local games
\cite{Oreshkov:2012aa}, in gate discrimination
\cite{PhysRevA.86.040301}, and oracle permut ation
\cite{COLNAGHI20122940,Facchini2015324}. This opened
the route to the study of operational tests for
indefinite causal structures based on the idea of
witnesses of a convex set
\cite{1367-2630-17-10-102001}, as well as to a notion
of dynamics of causal
structures~\cite{PhysRevX.8.011047}. The theoretical
effort in this field inspired pioneering experiments
\cite{Procopio:2015ab,PhysRevA.93.052321}.

The wealth of theoretical results about special cases of
 {higher-order quantum maps} calls for a thorough
unified theoretical
framework. This was initiated in Ref.~\cite{Perinotti2017}
and formalized in Ref.~\cite{8005095} in the language of
categorical quantum mechanics
\cite{Abramsky:1994ab,coecke2017picturing}. In the present
paper, we complete the picture with a fully operational
formulation. Every approach so far postulates complete
positivity as a purely mathematical requirement on
higher-order maps. Here we make the definition of
admissibility fully operational, avoiding explicit reference
to the mathematical properties of maps in the hierarchy---in
particular complete positivity is not postulated but
derived---and provide a characterization of admissible maps
thus defined.  {Higher-order quantum theory} must be
thought of as an extension of quantum theory, which provides
a natural unfolding of a part of the theory that is
implicitly contained in any of its formulations. As such,
 {it} has a fundamental value, being a new standpoint for
the analysis of the peculiarities of quantum theory. The
formulation of the theory of higher-order maps in terms of
operational axioms can indeed be applied to any operational
probabilistic theory---taking in due care the fact that in
general theories the notion of a transformation is more
complex
\cite{PhysRevA.81.062348,chiribella2011informational,d2017quantum}---and
allows for a comparison between the extended structures thus
obtained.

The study of the hierarchy of higher-order maps requires a
formal language that accounts for all the kinds of maps that
can be defined. Following Ref. \cite{Perinotti2017} we
define a {\em type system} for higher-order maps. Every map
comes then with a type, which summarizes basic information
such as its domain and its range. For example, provided that
elementary types such as $A,B$ denote the sets of states of
elementary systems, the type $(A\to B)$ denotes the set of
quantum operations with input is $A$ and output $B$.

 {Let us conclude this section with a short summary of
  the paper.  After a review of preliminary linear algebra
  and the Choi isomorphism in
  Sec.~\ref{sec:linear-maps-choi}, the type system of
  higher-order quantum maps is reviewed in
  Sec.~\ref{sec:types}, where the notion of extension by an
  elementary type is introduced, which plays a crucial role
  in the definition of admissibility. In
  Sec.~\ref{sec:axiom} the operational axioms of
  higher-order quantum theory are presented. We show that
  the property of complete positivity follows from the
  operational definition of admissibility
  provided. Moreover, we prove a necessary and sufficient
  condition for admissible maps to be deterministic, that
  will be used in the subsequent analysis. In
  Sec.~\ref{sec:charact} we introduce the notion of a type
  structure, which summarizes the important features of a
  type. Then we prove a characterization theorem for
  deterministic admissible maps of an arbitrary type which
  makes explicit the results of the previous section.  We
  then apply the result to some remarkable special cases,
  such as the proof of the uncurryng rule and the spelling
  out of the definition of tensor product of types. We also
  introduce the hierarchy of generalized combs, and show
  some structural identities for this family of maps. In
  Sec.~\ref{sec:inverse} we pose the problem of inverting
  the characterization of deterministic types, namely, given
  a convex set of maps, finding, if any, the type to which
  it corresponds. Finally, Sec.~\ref{sec:conclu} we close
  with some comments and remarks.}

\section{Linear maps and the Choi isomorphism}\label{sec:linear-maps-choi}

Let us start with some notational remarks.  We denote
quantum systems with capital letters $A,B \dots , Z$ and the
corresponding Hilbert spaces with
$\hilb{H}_A, \hilb{H}_B, \dots , \hilb{H}_Z$.  Throughout
this paper we restrict ourselves to quantum systems with
finitely many degrees of freedom, i.e. finite dimensional
Hilbert spaces.  The dimension of a Hilbert space
$\hilb{H}_A$ is denoted by $d_A$ and since $d_A < \infty$ we
have $\hilb{H}_A \equiv \mathbb{C}^{d_A}$.  The system with
dimension $1$, called the \emph{trivial system}, is denoted
by $I$.  The parallel composition of systems $A$ and $B$ is
denoted by $ AB$ and therefore we have
$\hilb{H}_{AB} = \hilb{H}_{A} \otimes \hilb{H}_{B}$.  The
parallel composition between a system $A$ with the trivial
system $I$ gives back the same sytem $A$, i.e. $AI = A$.  We
denote with $\Lin{(\hilb{H}_A)}$ the set of linear operators
on $\hilb{H}_A$ and with
$\mathcal{L}(\mathcal{L}(\hilb{H}_A),
\mathcal{L}(\hilb{H}_B))$ the set of linear maps from
$\mathcal{L}(\hilb{H}_A)$ to $\mathcal{L}(\hilb{H}_B)$.

A \emph{state} of a quantum system $A$ is a positive
operator $ 0 \leq \rho \in \mathcal{L}(\hilb{H}_A)$ such
that $\Tr[\rho] \leq 1$.  States such that $\Tr[\rho] =1$
are called normalized states or deterministic states.
Physical transformations from system $A$ to $B$ are
described by completely positive trace non increasing maps
$\mathcal{M} \in \mathcal{L}(\mathcal{L}(\hilb{H}_A),
\mathcal{L}(\hilb{H}_B))$ also known as \emph{quantum
  operations}.  The requirements of complete positivity and
trace non increasing guarantee that the transformation
$\mathcal{M}$ is physically \emph{admissible}, i.e. i) it is
compatible with the probabilistic structure of quantum
theory, and ii) it maps quantum states to quantum states
even when locally applied to bipartite states.  A quantum
operation which is trace preserving is called \emph{quantum
  channel}.  A set $\{ \mathcal{M}_i\}_{i\in \set S}$ of
quantum operations from system $A$ to system $B$ such that
$\mathcal{M} := \sum_{i \in \set S} \mathcal{M}_i $ is trace
preserving, is called \emph{quantum instrument}.  A special
instance of instrument is given by positive-operator-values
measures POVMs, which maps states into probabilities, and
are described by a collection of positive operators that
sums to the identity.  Moreover, states of a quantum system
$A$ can be considered as a special case of completely
positive maps from the trivial system $I$ to $A$.

The Choi isomorphism \cite{CHOI1975285} between linear maps
and linear operators will play a key role in the following.
\begin{theorem}[Choi isomorphism]\label{thm:choi}
  Consider the map $\choi:\mathcal{L}(\mathcal{L}(\hilb{H}_A),\mathcal{L}(\hilb{H}_B))
\to \mathcal{L}(\hilb{H}_B \otimes  \hilb{H}_A)  $ defined as
\begin{align}
  \choi:\mathcal{M} \mapsto M \qquad
  M:=\mathcal{I}_{A}\otimes  \mathcal{M}  (\KetBra{I}{I})
\end{align}
where $\mathcal{I}_{A}$ is the identity map on
$\mathcal{L}(\hilb{H}_A)$ and
$\Ket{I}:=\sum_{n=1}^{d_A} \ket{n}\ket{n}$,
$\{ \ket{n} \}_{n=1}^{d_A}$ denoting an orthonormal basis of
$\hilb{H}_A$.  Then $\choi$ defines an isomorphism between
$\mathcal{L}(\mathcal{L}(\hilb{H}_A),\mathcal{L}(\hilb{H}_B))$
and $\mathcal{L}(\hilb{H}_A \otimes \hilb{H}_B)$.  The
operator $M = \choi(\mathcal{M})$ is called the \emph{Choi
  operator} of $\mathcal{M}$. Moreover one has
\footnote{$\Tr_B$ denotes the partial trace on system $B$
  and $I_A$ is the identity operator on system $A$}:
\begin{align*}
 &\Tr[\mathcal{M}(X)] = \Tr[X] \quad \forall X \in \mathcal{L}(\hilb{H}_A)
\,
\Leftrightarrow
\,
   \Tr_B[M] = I_A ,
  \\
  &\mathcal{M}(X)^\dagger =  \mathcal{M}(X^\dagger)
  \,
  \Leftrightarrow
  \,
M^\dagger = M,\\
&  \mathcal{M}
  \mbox{ is completely positive}
  \;
\Leftrightarrow
\;
 M \geq 0.  
\end{align*}
\end{theorem}
The inverse of the map $\choi$ is given
by the following expression:
\begin{align}
  \label{eq:invchoi}
 & [\choi^{-1}(M)](O) = \Tr_A[(O^T\otimes I_B)M] \\
  &O \in \mathcal{L}(\mathcal{H}_A) \quad M\in
  \mathcal{L}(\mathcal{H}_A\otimes \mathcal{H}_B),
  \nonumber
\end{align}
where $O^T$ denotes the transpose opearator
with respect to the orthonormal basis
we used to define $\Ket{I}$ in Theorem \ref{thm:choi}.

\section{Type system}\label{sec:types}

In this section we lay the foundations of higher-order
quantum theory. The notions of quantum operation and POVM
allow for a complete and effective description of
processing of quantum information encoded into quantum states. 
However, this set of tools is unsuitable for describing
processes in which the input and output of the transformation are
transformations themselves. Our goal is to introduce a formal language
which enables us to overcome such a limitation.
This language can be regarded as the \emph{type system}
for  {higher-order quantum maps}. Starting with a set of \emph{elementary
  types}, corresponding to finite dimensional quantum systems,
by using appropriate \emph{type constructors} one recursively builds
new types from old ones.
This procedure generates the whole hierarchy of types of admissible quantum
maps, which maps from quantum trasformations to
quantum transformations are a special case of. 

\begin{definition} [Types]\label{def:quantum-types}
  Every finite dimensional quantum system corresponds to a \emph{Type}
  $A$.  The elementary type corresponding to the tensor product of
  quantum systems $A$ and $B$ is denoted with $AB$.  The type of the
  \emph{trivial system} is denoted by $I$.  We denote with
  $\ETy$ the set of elementary types.  Let
  $\set{A} := \ETy \, \cup \, \{(\}\, \cup \,\{)\}\, \cup \,
  \{\to\}$ be an alphabet.  We define the \emph{set of types} as the
  smallest subset $\set{Types} \subset \set{A}^*$ such
  that\footnote{Please note that $\set{A}^*$ stands for the set of
    words of the alphabet $\set{A}$}
    \begin{itemize}
    \item $\ETy \subset  \set{Types}$,
      \item if $x,y \in   \set{Types}      $  then $(x\to y) \in  \set{Types}$.
      \end{itemize}
    \end{definition}  
  As one can easily verify, a type $x$ is given a by a string like
  $x=(((A_1\to A_2)\to (A_3\to A_1 ))\to (A_4 \to A_1) )$ where $A_i$
  are elementary types.  {According to the above definition, for every 
  pair of types $x$ and $y$, one can form a new type $(x\to y)$, where $x$ is the 
  tail (input) and $y$ the head (output) of an arrow. The new type $(x\to y)$ must be
  thought of as a new single entity that can be the head or the tail of a further arrow.}
  In order to lighten the notation, the
  outermost parentheses are usually omitted.
  As it will be clear soon, if $A,B,C$ and $D$ are elementary types, then the type
  $(A \to B) $ is the type of maps from system $A$ to system $B$ and
  the type  $(A \to B) \to (C \to D)$ is the type of maps from
  ``maps from $A$ to $B$'' to  ``maps from $C$ to $D$''.
  It is worth noticing that, for each type $x$
  there exist a positive integer $n$, $n$ types $x_i$ and an
  elementary type $A$ such that
  \begin{align}
    \label{eq:normaltype}
    x = x_1 \to (x_2 \to (x_3 \to \cdots ( x_n \to A))\cdots)
  \end{align}
The following definition will allow us to extend the notion of
admissible map to the whole hierarchy.

\begin{definition} [Extension with an elementary type]\label{dfn:extension}
Let $x \in \set{Types}  $ be a type and $E \in \ETy$ be an
elementary type.
The \emph{extension} $x \para E$ of $x$ by the elementary type $E$
is defined recursively as follows:
\begin{itemize}
\item for any $A, E \in \ETy$ we have
  $A \para E := AE $;
  \item for any $x,y \in \set{Types}$, 
$(x \to y) \para E := (x \to y \para E)$. 
\end{itemize}

\end{definition}
From the first item of definition \ref{dfn:extension} we see that the
parallel composition of elementary events is recovered.
From the
recursive definition, it is immediate to compute the parallel
composition $x \para E$ when $x$ is given explicitly. For example we
have:
\begin{align*}
  &(((A_1\to A_2)\to (A_3\to A_1 ))\to (A_4 \to A_1) ) \para E = \\
  &(((A_1\to A_2)\to (A_3\to A_1 ))\to (A_4 \to A_1)\para E) = \\
  &(((A_1\to A_2)\to (A_3\to A_1 ))\to (A_4 \to A_1\para E) )=\\
  &(((A_1\to A_2)\to (A_3\to A_1 ))\to (A_4 \to A_1 E) )   
\end{align*}
From Equation~\eqref{eq:normaltype} we clearly have
$ x \para E = x_1 \to (x_2 \to \cdots ( x_n \to AE)\cdots)$.
Clearly, the parallel composition with the trivial type $I$,
leaves the type $x$ unaffected, i.e. $x \para I = x$.
Since 
many of the results of this paper are proved by
induction, it is useful to introduce the following partial
ordering between types.
\begin{definition}[Partial ordering $ \preceq $]
  We say that type $x$ \emph{is a parent} of type $y$
  and we write $x \preceq_p y$
  if there exists a type $z$ such that
  either $y = (x \to z) $ or $y = (z \to x) $.
  The relation $x \preceq y$
  is defined as the transitive closure of the binary relation
  $\preceq_p$
\end{definition}
From the previous definition we have, for example,
\begin{align*}
  x = (y \to w) \to z \implies  y , w , z \preceq x.
\end{align*}
The relation $\preceq $ is a
well founded relation and Noetherian induction can be used.
If we want to show that some proposition $\mathfrak{P}(x)$ holds for all types $x$ of the set $\set{Types}$,
we need to show that:
\begin{itemize}
\item[1] $\mathfrak{P}(y)$ is true for all elementary types (which are the minimal elements of the
  set $\set{Types}$).
  \item[2] If $ \mathfrak{P}(y)$ is true for all $y$ such that $y \preceq x$, then
    $\mathfrak{P}(x)$
    is true for $x$.
  \end{itemize}
In most of the cases, we will be required to prove that a statement
holds for the type $x || E$ for any arbitrary elementary type $E$.
Then item $2$ becomes:
\begin{itemize}
  \item[2'] If $ \mathfrak{P}(y \para E)$ is true for all $y$ such that $y
    \preceq x$ and for any $E$, then
    $\mathfrak{P}(x \para E' )$
    is true for $x$ and any $E'$.
  \end{itemize}

\section{Axioms for higher-order quantum theory}\label{sec:axiom}
  
  It is worth stressing that the hierarchy of types
  has been defined as an abstract set of strings,
with no relationship with the set of linear maps on Hilbert space. We now introduce 
such a connection through the notion of event.
\begin{definition}[Generalized events] \label{def:eventsdef}
  If  $x$ is a type in $\Ty$, the set of \emph{generalized events of type $x$}, denoted by $\Ev x$,
  is defined by the following recursive definition.
  \begin{itemize}
  \item if $A$ is an elementary type, then every
    $M \in \mathcal{L}(\hilb{H}_A)$  is a generalized event of type $A$,
    i.e. $\Ev A := \mathcal{L}(\hilb{H}_A) $.
  \item if $x, y$ are two types, then every Choi operator of 
    linear maps $\mathcal{M} : \Ev x \to \Ev y$, is a generalized event  ${M}$ of type $(x \to y)$.
  \end{itemize}
\end{definition}
The following lemma immediately follows 
from of Definition~\ref{def:eventsdef}.
\begin{lemma}[Characterization of events]\label{lmm:choigeneric}
  Let $x$ be a type.
  Then   $\Ev x = \mathcal{L}(\hilb{H}_x) $
  where $ \hilb{H}_x := \bigotimes_{i}\hilb{H}_i $ and
$\hilb{H}_i$ are the Hilbert spaces corresponding to the
  elementary types $\{ A_i\}$ occuring in the expression of $x$.
\end{lemma}

\Proof
First we notice that the thesis holds for elementary types $x=A$. 
We then prove that if the thesis holds for
arbitrary types $x , y$ than it holds for $x \to y$.
Let us then suppose that $\Ev x = \mathcal{L}(\hilb{H}_x)$ and
$\Ev y = \mathcal{L}(\hilb{H}_y)$.  An event of type
$\Ev {x \to y}$ is the Choi operator $M$ of a map
$ \mathcal{M} : \mathcal{L}(\hilb{H}_x) \to \mathcal{L}(\hilb{H}_y)$
and therefore $M \in \mathcal{L}(\hilb{H}_x \otimes \hilb{H}_y)$.\qed


An explicit example can be useful.  Let $A,B,C$ and $D$ be elementary
quantum systems with Hilbert spaces
$\hilb{H}_A, \hilb{H}_B, \hilb{H}_C, \hilb{H}_D$ and let us consider
the type $x := ((A \to B) \to C)\to D$. According to Definition
\ref{def:eventsdef} and Lemma \ref{lmm:choigeneric}, an event of type
$x$ is an operator
$M \in \mathcal{L}(\hilb{H}_A \otimes \hilb{H}_B \otimes \hilb{H}_C
\otimes \hilb{H}_D)$.
Obviously, the type of an event cannot be inferred by the operator
alone. Indeed, the same $M$ can also define an event of a different type
$y := (A \to B) \to (C\to D)$. Therefore, when we define an event, we
need to explicitly declare its type.

Given two quantum systems $A$ and $B$,
not every operator $M\in \mathcal{L}(\hilb{H}_A \otimes \hilb{H}_B)$
represents the Choi operator of a physical transformation from $A$ to $B$.
An operator $M$ represents a physical transformation
if and only of it is the Choi operator of a completely positive trace non
increasing map, i.e. if and only if $ 0\leq M \leq N$
with $\Tr_B[N] = I$.
In an analogous way, we now want to characterise those events
that correspond to physical maps. The key step toward achieving this
goal is to formulate a notion of \emph{admissible event}
which generalises the requirement of complete positivity.
In order to do that, we start with the following definition.
\begin{definition}[Extended event]
  Let $x$ be a non-elementary type,
  $E$ an elementary  type
  and $M \in \Ev x$.
  We denote with $M_E$ the \emph{extension of $M$ by $E$}
  which is defined recursively as follows:
If $x,y$ are two types and
    $M \in \Ev {x\to y}$ then
    $M_E \in \Ev {x\para E \to y \para E}$
    is the Choi operator of the map
    $\mathcal{M} \otimes \mathcal{I}_E :
    \Ev {x\para E} \to   \Ev  {y \para E} $, where
     $\mathcal{I}_E : \mathcal{L}(\mathcal{H}_E) \to
     \mathcal{L}(\mathcal{H}_E)  $
     is the identity map.
\end{definition}
If $A$ and $B$ are elementary types
then  $M \in \Ev {A \to B}$
is the Choi of a map
$ \mathcal{M} : \mathcal{L}(\mathcal{H}_A) \to \mathcal{L}(\mathcal{H}_B) $.
Therefore, $M_E$
is the Choi operator of the map
$ \mathcal{M} \otimes \mathcal{I}_E:
\mathcal{L}(\mathcal{H}_A \otimes \mathcal{H}_E)
\to
\mathcal{L}(\mathcal{H}_B \otimes \mathcal{H}_E) $.

The notion of extended event allows us to give the definition of
admissible event. We split the definition into two parts.  The first
part defines admissible elementary events and it is
the usual definition of quantum states as positive operators.
\begin{definition}[Admissible elementary event] \label{def:admelemev}
      Let $A$ be an elementary type and $M \in \Ev A$.
We say that:
\begin{itemize}
\item  $M$ is a \emph{deterministic}  event if
$M \geq 0$ and  $\Tr[M]=1$. $\Evd A$ denotes the set of deterministic events of type $A$. 
\item $M$ is \emph{admissible} 
if $M \geq 0$ and there exists   $N \in \Evd A$ such that
$M \leq N$.
  $\Eva A$ is the set of admissible events of type $A$.
\end{itemize}
\end{definition}
Admissible elementary events which are not deterministic, i.e. the strict
inequality
$M < N$ holds, are called \emph{probabilistic} elementary events.
Up to this point, Definition~\ref{def:admelemev} just introduced a new notation for well known objects.
However, the use of this new language simplifies the statement of the
second part of the definition of admissible events.
\begin{definition}[Admissible event]\label{def:admissiblevents}
Let $x,y\in\Ty$ be two types, $M \in \Ev {x\to y}$
be  an event of type $x\to y$ and 
$M_E \in \Ev {x\para E\to y \para E}$ be
the extension of $M$ by $E$.
Let  $\mathcal{M} : \Ev x \to \Ev y $
and  $\mathcal{M} \otimes \mathcal{I}_E :
  \Ev {x \para E} \to \Ev {y \para E} $ be
the linear maps whose Choi operator are $M$
and $M_E$ respectively.

We say that $M$ is \emph{admissible}  if,

\begin{enumerate}
\item \label{item1}for all elementary types $E$, the map
  $\mathcal{M} \otimes \mathcal{I}_E$ sends
  admissible events of type $x \para E$
  to
   admissible events of type $y \para E$.
\item there exist $\{N_i\}_{i=1}^n  \subseteq\Ev {x\to y} $, $0\leq n<\infty$
such that, for all elementary types $E$,

\begin{itemize}
\item[$\ast$] $\forall 1\leq i\leq n$ The map $\mathcal{N}_i$ satisfies item \ref{item1}, 
\item[$\ast$] For all elementary types $E$, the map $(\mathcal{M} + \sum_{i=1}^n\mathcal{N}_i ) \otimes \mathcal{I}_E  $ maps deterministic events of type $x\para E$ to deterministic events of type $y\para E$
\end{itemize}
\end{enumerate}

The set of admissible events of type 
$x \to y$ is denoted with $ \Eva {x\to y}$.
An operator  $D \in \Ev {x\to y}$ is 
a  \emph{deterministic}
event of type $x \to y$, if  $D \in \Eva {x\to y}$
and $(\mathcal{D} \otimes \mathcal{I}_E)$ maps deterministic admissible events of type $x \para E$ to deterministic admissible events of type $y\para E$.
\end{definition}
In lemma~\ref{lem:sticazzetti} we prove that
if $M \in  \Eva {x}$,
and $\{N_i\}_{i=1}^n$ satisfy the requirements
of Definition \ref{def:admissiblevents},
then $N_i \in  \Eva {x}$
and $M+\sum_{i=1}^n N_i \in  \Evd {x}$.
Clearly, we also have that if, for every $E\in\ETy$,
$\mathcal{D}\otimes \mathcal{I}_E (\Eva {x \para E} )
\subseteq
\Eva {y \para E}$
and
$\mathcal{D} (\Evd {x \para E } )
\subseteq
\Evd {y \para E }$,
then $D \in \Evd {x \to y}$.

 {Definition \ref{def:admissiblevents} generalises Kraus' axiomatic 
  definition of quantum operations \cite{Kraus} to higher-order maps.
  Indeed, one can easily verify that, for the simplest case $x = A \to B$,
  definition \ref{def:admissiblevents} reduces to the notion of
  completely positive trace non increasing map from
  $\mathcal{L}(\hilb{H}_A)$ to $\mathcal{L}(\hilb{H}_B)$.  Let $M$ be
  an admissible event of type $A \to B$.  Since the set of admissible
  event of type $A$ and $B$ are the set of density matrices,
  $\mathcal{M}$ must be completely positive.  Moreover, there 
  exists a set of operators $N_i$ such that, for any $i$,
  $\mathcal{N}_i$ must be completely positive as well.
  The condition that  $\mathcal{M} +\sum_i \mathcal{N}_i$ maps
  deterministic events of $A$ to deterministic events of $B$,
  implies that $\mathcal{M}
  +\sum_i \mathcal{N}_i$ must be trace preserving and therefore
  $\mathcal{M}$ is trace non increasing.
}

The following theorems characterise the set of admissible events.
\begin{theorem}[Characterization of admissible
  events]\label{prop:positiveconeadmis}
  Let $x$ be a type and $M\in\Ev x$.
  Then we have
  \begin{align}
    &M \in \Eva {x}\ \Leftrightarrow\ 
      M \geq 0  \; \land \; \exists D \in
    \Evd {x} \mbox{ s.t. } M \leq D ,
\label{eq:positivity}    
  \end{align}
\end{theorem}  
\Proof
See Appendix~\ref{sec:proof-theor-refpr}.
\qed

The result of theorem \ref{prop:positiveconeadmis} tells us that the
only relevant cone in higher-order quantum theory is the cone of
positive operators.  This is a relevant improvement e.g.~with respect
to the previous literature on the subject, where complete positivity
was assumed from the very beginning. The present definition of
admissibility, on the contrary, can be extended to the case of general
operational probabilistic theories \cite{d2017quantum,PhysRevA.81.062348,d2014feynman}
where in general
the Choi correspondence, defined through the notion of a {\em faithful
  state}, is not surjective on the cone of states.

Notice that condition \eqref{eq:positivity} reduces the characterization of the set $\Eva x$ to that 
of the set of deterministic events $\Evd x$. The latter is achieved by the next result.

\begin{theorem}[Characterization of deterministic  events]
  \label{prop:deteventmadeeasy}
Let $x,y$ be two types, $M \in \Ev {x\to y} $
be  an event of type $x \to y$.
Then we have:
  \begin{align}
    \label{eq:charadetprelimin}
    M \in \Evd {x \to y} \! \iff
   \! \! \left 
    \{ \!  
    \begin{array}{l}
    M \geq 0, \\
  {[}\Ch^{-1}(M) {]}  (\Evd{x}) \subseteq  \Evd{y}
    \end{array}
\right.
  \end{align}
\end{theorem}
\Proof
See Appendix~\ref{sec:proof-prop-deteasy}. 
\qed

$\,$

Definition~\ref{def:admissiblevents},
Theorem~\ref{prop:positiveconeadmis}
and
Theorem~\ref{prop:deteventmadeeasy}
complete the construction of the
hierarchy of higher-order quantum maps. Every type $x$ corresponds
to a convex set of positive operators which is the set
$\Evd {x}$ of deterministic events of type $x$.  The set
$\Evd {x}$ uniquely determines the convex set $\set{PType}_x$
of probabilistic events of type $x$.
According to our framework,
the colloquial sentence
``$M$ is an higher-order quantum map'' translates into
``$M$ is a deterministic or probabilistic event of some kind $x$''

The main question in the theory of
 {higher-order quantum theory} is to characterize $\Evd {x}$
for any type $x$. For example, one could ask whether two different
different types $x$ and $y$ have the same set of deterministic events,
i.e.  $\Evd {x} = \Evd {y}$.  Whenever this is the case,
we say that the types $x$ and $y$ are \emph{equivalent}. We emphasize
this concept by giving the following definition.
\begin{definition}[Equivalent types]
  Let $x$ and $y$ be two types. We say that
  $x$ and $y$ are \emph{equivalent}, and denote it as $x \equiv y$,
  if $\Evd {x} = \Evd {y}$.
  \label{def:equivtyp}
\end{definition}

\section{Characterization of higher-order quantum maps}\label{sec:charact}

In this section we further develop the  { framework of
higher-order quantum
theory} that has been introduced in the previous section.
\subsection{Type structure}
Many
results we are going to prove depend only on the structure of the type
$x$ we are considering rather than on the specific elementary systems $A_i$
that compose it. For example,
the types $A_0 \to B_0$ and $A_1 \to B_1$
will be treated on the same footing, even if $d_{A_0} \neq d_{A_1}$
or $d_{B_0} \neq d_{B_1}$. 
It is then convenient to give the following definition.
\begin{definition}[Type structure]
  Let
  $\set{\Omega} $ $:= $ $\{*, I,(,),\to\}$ be an alphabet.  We define the \emph{set
    of type structures} as the smallest subset
  $\set{Str} \subset \set{\Omega}^*$ such that
    \begin{itemize}
    \item $* \, ,  I\in  \set{Str}$,
    \item if $\mathrm{x}, \mathrm{y} \in   \set{Str}      $
      then $(\mathrm{x} \to \mathrm{y}) \in  \set{Str}$.
    \end{itemize}
    We say that a type $x$ belongs to the type structure $\mathrm{x}$,
    and we write
$x \in \mathrm{x} $, if $x$ can be obtained 
by substituting arbitrary elementary types $A_i\in\ETy$ (that can possibly be the trivial
type $I$) in place of the symbols $*$ in the expression of the
type structure $\mathrm{x}$.
\end{definition}
One could think of  a structure as an expression of the kind
\begin{align}
  \label{eq:structureform}
  \mathrm{x} :=
  (( * \to *) \to I)\to( * \to *),
\end{align}
and the types that belong to $\mathrm{x}$ are, for example,
\begin{align*}
  ((A \to B) \to I)\to( C \to D) \in  \mathrm{x} \\
  ((A \to I) \to I)\to( I \to D) \in  \mathrm{x} 
\end{align*}
The type structure $\mathrm E$ is the type structure of the elementary
types, $A \in \mathrm E$ $\forall A \in \ETy$. Given a type structure
$\mathrm{y}$ one can obtain another type structure $\mathrm{y}$' by substituting the
trivial type $I$ in place of some of the symbols
$*$ in the expression of $\mathrm{y}$.
This feature introduces a
partial ordering among the type structures:
\begin{definition}[Substructures]
  We say that a type strucure $\mathrm{x} $
  is \emph{substructure} of a  type strucure $\mathrm{x}' $ ad we
  write
  $\mathrm{x}  \subset \mathrm{x}' $
  if  $\mathrm{x} $ can be obtained 
   by substituting the
trivial type $I$ in place of some of the symbols
$*$ in the expression of $ \mathrm{x}' $.
\end{definition}
For example we have:
\begin{align*}
\mathrm{y}& :=   (* \to I  ) \to( *  \to
               * )  \\
  \mathrm{y}' &:=   (* \to *  ) \to( *  \to
               * )  \\
\mathrm{y}  & \subset \mathrm{y}'.
\end{align*}

We notice that the same type $x$ may belong to different type
strucures, for example
\begin{align*}
  \mathrm{y} &:=   (* \to *  ) \to( *  \to
               * )  \\
  \mathrm{y}'& :=   (* \to I  ) \to( *  \to
               * )  \\
  y &:=  (A \to I) \to( C \to D) \quad y \in \mathrm{y} , \mbox{ and }y \in \mathrm{y}'.
\end{align*}
However, among the type structures which a type $x$ belongs to, there
exists a privileged one.
\begin{definition}[Natural type structure]\label{def:nattypestruct}
  The \emph{natural type structure}  of a type $x$,
  is the type structure $[x]$ such that:
  \begin{itemize}
  \item $x \in [x]$
    \item $x \not \in \mathrm{y}$ for any $\mathrm{y} \subset [x]$
  \end{itemize}
\end{definition}
The expression of the natural type structure of a type $x$ is obtained
by replacing  the all the elementary types but the trivial ones in the
expression of $x$, with
the elementary type structure $*$.
The following example clarifies the meaning of Definition~\ref{def:nattypestruct}:
\begin{align*}
  x :=  (A \to I) \to (( C \to D) \to (F \to I) )\\
  [x] :=
  (* \to I) \to (( * \to *) \to (* \to I) ).
\end{align*}



\subsection{$\set{L}_{\bvec{b}} $ spaces}\label{sec:setl_bvecb--spaces}
There is family of linear spaces of operators
that plays a central role
in  { higher-order quantum theory.}
In this subsection, we will introduce a notation which will allow
us to more efficiently manipulate those linear spaces.

For a given an Hilbert space $\hilb{H}$,
we denote with $\set{Herm}(\hilb{H})$ the
the linear (real) subspace  
of the Hermitian operators on $\hilb{H}$.
It is useful to split $\set{Herm}(\hilb{H})$
as the direct sum of the subspace of traceless operators and the one
dimensional subspace generated by the identity operator:
\begin{align}
  \label{eq:tracelessbasis}
  \begin{aligned}
 & \set{L}_1 := \spn\{ I \} \quad \set{L}_0 := \{ X \mid  \Tr[X] = 0,\ X^\dag=X  \}\\
      &\set{Herm}(\hilb{H})=\set L_0\oplus\set L_1
  \end{aligned}
\end{align}
where $I$ is the
identity operator on $\hilb{H}$.  Therefore, if $O$ is in
$\set{Herm}(\hilb{H})$ we can write the decomposition
$O = \lambda I + X$ where $\lambda \in \mathbb{R}$ and $T$ is a
traceless selfadjoint operator $X \in \set{L}_0$, $X $.  When we are dealing with a tensor product of
$l$ Hilbert spaces,
$\hilb{H} := \hilb{H}_{A_1} \otimes \hilb{H}_{A_{2}} \otimes \dots
\otimes \hilb{H}_{A_l} $, we define
\begin{align}
  \label{eq:tracelessbasis2}
  \begin{aligned}
     \set{L}_{\bvec{b}} := \set{L}_{b_1} \otimes  \set{L}_{b_{2}}
     \otimes \dots \otimes  \set{L}_{b_l}  \quad b_i=0,1.
  \end{aligned}
\end{align}
For example, for $\hilb{H} := \hilb{H}_{A_{1}}  \otimes \hilb{H}_{A_2} $,
we have
\begin{align*}
  \begin{aligned}
&  \set{L}_{00} = \spn\{ X \otimes Y \} \quad
  \set{L}_{01} = \spn\{ X \otimes I   \} \\
  &\set{L}_{10} = \spn\{ I \otimes Y  \} \quad
  \set{L}_{11} = \spn\{ I \otimes I   \}, 
  \end{aligned}
\end{align*}
where the symbols $X$ and $Y$ denote $X\in\set L_0$ and $Y\in\set L_0$, respectively.

  It is rather easy to verify that the spaces
  $\set{L}_{\bvec{b}} $ enjoy  the following properties:
  \begin{lemma}[Properties of the $ \set{L}_{\bvec{b}} $ spaces]
    \label{lmm:Lbspaces}
    Given a binary string $\bvec{b}$  of lenght $l$,
    let $ \set{L}_{\bvec{b}}$ be the corresponding
    subset of $\hilb{H} := \hilb{H}_{A_1} \otimes \hilb{H}_{A_{2}} \otimes
    \dots \otimes \hilb{H}_{l} $ defined as in Equation
    \eqref{eq:tracelessbasis2}.  If $\bvec{b} \neq \bvec{b'}$
      then $ \set{L}_{\bvec{b}}$ and $ \set{L}_{\bvec{b'}}$
      are orthogonal subspaces with respect the Hilbert-Schmidt
      product\footnote{we remind that the Hilbert-Schimt product of
        two operators $A$ and $B$ is $(A,B)_{HS}:= \Tr[A^\dag B]$}.
  \end{lemma}
\Proof Since
$\bvec{b} \neq \bvec{b}'$
there exist an some $i$
such that
$b_i \neq b'_i$.
Without loss of generality we may suppose
that
$b_1=1 $ and $b'_1=0$.
From Equation \eqref{eq:tracelessbasis2} we have
$ \set{L}_{\bvec{b}} \ni A=  I\otimes \tilde{A} $
and $ \set{L}_{\bvec{b}'}\ni B = X\otimes\tilde{B}$.
Taking the Hilbert-Schmidt product of $A$ and $B$ gives
\begin{align*}
  (A,B)_{HS} &= \Tr[A^\dag B]=\Tr[(I\otimes \tilde{A}^\dag )(X\otimes \tilde{B}
  )]\\
  &= \Tr[X]\Tr[\tilde{A}^\dag \tilde{B}] = 0.
\end{align*}
This proves that
$\set{L}_{\bvec{b}}$ and $ \set{L}_{\bvec{b}'}$
are Hilbert-Schmidt orthogonal. \qed

\noindent From Lemma \ref{lmm:Lbspaces}
we have that
the sum $\set{L}_{\bvec{b}} + \set{L}_{\bvec{b}'}$
is actually a direct sum
$\set{L}_{\bvec{b}} \oplus \set{L}_{\bvec{b}'}$.
It is useful to intruduce the following notation:
\begin{align}
  \label{eq:tracelessspace}
  & W^{(l)}:= \{0,1\}^l, \quad  T^{(l)} :=W^{(l)} \setminus \{\bvec{e}\}, \;
     \bvec{e} := 11\dots 1, \\
  &\set L_J:=\bigoplus_{\bvec b\in J} \set L_{\bvec b},\quad J\subseteq W^{(l)},\quad \set L_\emptyset=\{0\},\ \set L_{\boldsymbol{\varepsilon}}=\mathbb R,
\end{align}
where ${\boldsymbol{\varepsilon}}$ is the null string in $W^{(0)}$ such that
\begin{align}
{\boldsymbol{\varepsilon}}\bvec b=\bvec b{\boldsymbol{\varepsilon}}=\bvec b\quad\forall \bvec b\in W^{(l)}.
\end{align}

It is worth stressing that the notation $ \set{L}_{\bvec{b}}$ is not
  reminiscent of the dimensions of the Hilbert spaces
  $\hilb{H}_{A_i}$ occurring in the decomposition $\hilb{H} = \bigotimes_i{\hilb{H}_{A_i}}$.  Therefore, if two types have the same natural type structure, i.e.
  $[x]  = [x']$,   they share the same set of strings.

Given a subspace
\begin{align*}
  &\set{\Delta}  = \bigoplus_{\bvec{b} \in J \subseteq T^{(l)}}
  \set{L}_{\bvec{b}},
\end{align*}
we define the following two spaces related to $\set \Delta$
\begin{align}
  &\overline{\set{\Delta}}:= \bigoplus_{\bvec{b} \in \overline{J}}
    \set{L}_{\bvec{b}}, && \overline{J} := T^{(l)} \setminus J.
    \label{eq:tracelesspace2}\\
    &{\set{\Delta}}^\perp:= \bigoplus_{\bvec{b} \in {J}^\perp}
    \set{L}_{\bvec{b}}, && {J}^\perp := W^{(l)} \setminus J.
    \label{eq:compspace}
\end{align}
Given $J\subseteq W, J'\subseteq W^{(l')}$ and $\bvec w'\in W^{(l')}$, we can define the sets $JJ'\subseteq W^{(l+l')}$ and $J'\bvec w'\subseteq W^{(l+l')}$
as follows:
\begin{align}
  \begin{split}
  \label{eq:concatset}
  &JJ':=\{ \bvec{b}=\bvec{w} \bvec{w'} \, | \, \bvec{w}\in J ,\, \bvec{w'}\in J'   \},\\
  &J\bvec w':=\{\bvec w\bvec w'\mid\bvec w\in J\}.
  \end{split}
\end{align}
If $J=J'$ we will write $J^2=JJ$ and $J^n$ for the set $JJ... $ (n times).
In the following we will omit the label $l$ from the symbols $W^{(l)}$
and $T^{(l)}$, whenever $l$ is clear from the context.
Eq. \eqref{eq:tracelesspace2} defines the complement of $\set{\Delta}$
in the space of traceless operators, i.e. $\set{\Delta} \oplus
\overline{\set{\Delta}} = \set{Traceless}(\hilb{H})$. Notice that, according to the definitions above, we have
\begin{align}
&\overline{\set L}_J=\set L_{\overline J},\quad J\subseteq T\\
&\set{Herm}(\hilb{H})= \set L_{W}=\bigoplus_{\bvec{b} \in W }
  \set{L}_{\bvec{b}},\\
&\set{Traceless}(\hilb{H})= \set L_{T}=\bigoplus_{\bvec{b} \in T }
  \set{L}_{\bvec{b}}. 
\end{align}
Notice that when $\hilb H=\bigotimes_i\hilb H_{A_i}$ contains some trivial system $A_{k}=I$, one has $\hilb H=\bigotimes_{i\neq k}\hilb H_{A_i}$. Correspondingly, the non trivial spaces $\set L_\bvec b$ are determined only by the bits $b_j$ in positions $j\neq k$ corresponding to systems different from the trivial system $A_k$. Indeed, if $b_k=0$ the space $\set L_\bvec b=\{0\}$ is trivial, while for $b_k=1$ one has that $\set L_\bvec b=\set L_{\bvec b'_k}$, where $\bvec b'_k$ is the string obtained from $\bvec b$ dropping the $k$-th bit. In formula
\begin{align}
&\set L_J=\set L_{J'_k},\quad J'_k:=\{\bvec b'_k\mid\bvec b\in J,\ b_k=1\},\\
&\bvec b'_k:=b_1b_2\ldots b_{k-1}b_{k+1}\ldots b_l,
\end{align}
having denoted by $l$ the length of the string $\bvec b$ so that the strings $\bvec b'_k$ have length $l-1$.
Repeatedly reducing the expression of the space 
$\hilb H=\bigotimes_i\hilb H_{A_i}$ to $\hilb H=\bigotimes_{i\not\in N}\hilb H_{A_i}$, where $N:=\{i\mid A_i=I\}$, one obtains
\begin{align}
\begin{aligned}
&\set L_J=\set L_{J'_N},\quad J'_N:=\{\bvec b'_N\mid\bvec b\in J,\ \forall k\in N\, b_k=1\},\\
&\bvec b'_N:=((\bvec b'_{k_1})'_{k_2}\ldots)'_{k_n},\quad k_i\in N,\ n=|N|.
\end{aligned}
\label{eq:normform}
\end{align}
Once a set $J$ is reduced as above, dropping all the bits in positions $i\in N$ corresponding to trivial systems $A_i=I$, we call the resulting set of strings $J'_N$ to be reduced to its {\em normal form}. The strings in $J'_N$ have length $l-n$. Notice that for the trivial system $I$ with $\hilb H=\mathbb C$, we have $W=W^{(0)}=\{{\boldsymbol{\varepsilon}}\}$, and correspondingly
\begin{align}
&\set{Herm}(\hilb H)=\set L_{\boldsymbol{\varepsilon}}=\mathbb R,\\
&\set{Traceless}(\hilb H)=\set L_{\emptyset}=\{0\}.
\end{align}

\subsection{Characterization of $\Evd{x}$}

We are now ready to present the characterization of the
set $\Evd {x}$ of deterministic events of type $x$.
The first step is to prove the following lemma.
\begin{lemma}[Transpose of deterministic event]
  \label{lem:transposedetevent}
  Let $x$ be a type and let
  $R \in \Evd x$ be a deterministic event of type $x$.
  Then also $R^T$, which is the transpose of $R$ with respect to the basis
  used in the definition of the Choi isomorphism,
  is a deterministic event of type $x$, i.e.
  $R \in \Evd x \iff R^T \in \Evd x$.
\end{lemma}
\Proof
The statement is true for elementary events.
Let us suppose that the statement is true for arbitrary types
$x$ and $y$ and let $R$ be a deterministic event of type $x \to y$.
Then we have
$\Tr_x[(S^T_x \otimes I_y)R] \in \Evd y$ for any
$ S_x\in \Evd x$.
By hypothesis we have that
$S^T_x\in \Evd x$
for any
$ S_x\in \Evd x$,
and therefore
$\Tr_x[(S_x \otimes I_y)R] \in \Evd y$
for any
$ S_x\in \Evd x$.
By hypothesis we also have that
$S^T_y\in \Evd y$
for any
$ S_y\in \Evd y$,
and consequently
$\Tr_x[(S^T_x \otimes I_y )R^T]=(\Tr_x[(S_x \otimes I_y)R])^T  \in \Evd y$
which by theorem \ref{prop:deteventmadeeasy} proves
$R^T \in \Evd {x \to y}$. \qed





We now prove the main result of this section.
\begin{proposition}[Characterization of $\Evd x$]
  \label{prop:char-high-order}
  Let $x$ be a type and let $ \hilb{H}_x := \bigotimes_{i}\hilb{H}_i $ be
  the Hilbert space given by the tensor product of
  the Hilbert spaces corresponding to the
elementary types $\{ A_i\}$ occuring in the definition of $x$.
Then we have:
\begin{align}
  \label{eq:characterization}
  \begin{aligned}
&    R \in \Evd x \iff R = \lambda_x I_x + X_x\\
& X_x \in \set{\Delta}_x\subseteq  \set{Traceless}(\hilb{H}_x), \quad R \geq 0 ,
  \end{aligned}
\end{align}
where the real positive coefficient $\lambda_x$ and
and the linear subspace
$\set{\Delta}_x$ are defined recursively as follows:
\begin{align}
  \begin{split}
   \set{\Delta}_A &=  \set{Traceless}(\hilb{H}_A), \mbox{ if }A \in \ETy\\
       \set{\Delta}_{x \to y} &=
[\set{Herm}(\hilb{H}_x) \otimes {\set{\Delta}_{y}}]
    \oplus
[\overline{\set{\Delta}}_{x}   \otimes \set{\Delta}^\perp_{y} ],
         \label{eq:eledetchar2}    
       \end{split}
\\                                  
    \begin{split}
    \lambda_{E} &= \frac{1}{d_E} \mbox{ if }E \in \ETy ,
       \quad
    \lambda_{x \to y} = \frac{\lambda_y}{d_x \lambda_x}. 
  \end{split}
    \label{eq:lambdarecursive}
  \end{align}
  \end{proposition}
  \Proof
  For any elementary type $A$ the set
  $\Evd A$ is the set of normalized states, and then the thesis
  holds.
  Let us consider the case in which $x$ is not elementary
  and let us suppose that the thesis hold for any type $y \preceq x$.
  For $x = y \to z $, any $R \in \Evd {y \to z}$ is a positive operator that
  can be decomposed as $R = \lambda_R I + O_R$
  where $ O_R \in \set{Traceless}(\hilb{H}_{y} \otimes \hilb{H}_z)$. 
  Since $R$ maps deterministic events of type
  $y$ to deterministic events of type $z$,
  we must have
  $  \Tr_y[(S_y^T \otimes I_z) R ] \in \Evd {z}$
  for all $S_y \in \Evd {y}$.
  Thanks to Lemma \ref{lem:transposedetevent}, this can be restated as
$  \Tr_y[(S_y \otimes I_z) R ] \in \Evd {z}$
for all $S_y \in \Evd {y}$.
First, let us consider the case
$S_y = \lambda_y I_y$, which is in $\Evd {y}$ thanks to the inductive
hypothesis.
From the inductive hypothesis, there exists $Z \in \set{\Delta}_z$
such that
\begin{align}
  \label{eq: charactR}
  \Tr_y[(\lambda_y I_y \otimes I_z) (\lambda_R I + O_R )] = \lambda_zI_z
+ Z
\end{align}
Equation~\eqref{eq: charactR} implies
that
\begin{align}
  \lambda_R =\frac{\lambda_z}{d_y  \lambda_y}=: \lambda_{y \to z}=\lambda_x
  \label{eq:charactlambdaR} \\
  \Tr_y[ O_R ] \in \set{\Delta}_z.
  \label{eq:charactOR1}
\end{align}
Let now $Y$ be an arbitrary operator in $\set{\Delta}_y$.
There exists $\mu \neq 0$ such that
$\lambda_y I + \mu Y  \geq 0$. From the induction hypothesis we have
$\lambda_y I + \mu Y  \in \Evd{y}$, which implies, together with
Equation
\eqref{eq:charactOR1},
\begin{align}
  \label{eq:charaOR2}
  \Tr_y[((\lambda_y I_y + \mu Y) \otimes I_z)( \lambda_{x} I_x
  + O_R) ] = \lambda_zI_z + Z,
\end{align}
for some $Z \in \set{\Delta}_z$.
From Equations~\eqref{eq:charactlambdaR}~\eqref{eq:charactOR1} and~\eqref{eq:charaOR2}
we obtain that
\begin{align}
\label{eq: charaOR3}
    \Tr_y[ (X_y \otimes I_z) O_R ] \in \set{\Delta}_z, \quad \forall
  X_y \in \set{\Delta}_y.
\end{align}
Equations~\eqref{eq:charactOR1} and \eqref{eq: charaOR3}
are satisfied if and only if
\begin{align}
  \label{eq:charaOR4}
  \Tr[ (X_y \otimes \overline{S}_z) O_R ] = 
  \Tr[ (I_y \otimes \overline{S}_z) O_R ] = 0, 
\end{align}
for any
$X_y \in \set{\Delta}_y$
and any
$ \overline{S}_z \in \set{\Delta}_z^\perp $.
    Equation~\eqref{eq:charaOR4} finally implies 
    \begin{align}
       O_R \in
[\set{L}_W \otimes {\set{\Delta}_{z}}]
    \oplus
[\overline{\set{\Delta}}_{y}   \otimes \set{\Delta}_z^\perp  ].  
    \end{align}
On the other hand,
let us consider and arbitrary operator $O'_R \in [\set L_W \otimes {\set{\Delta}_{z}}]
    \oplus
[\overline{\set{\Delta}}_{y}   \otimes \set{\Delta}_z^\perp ]$.
Clearly, there exist a real number $\mu \in \mathbb{R}$
such that $R' := \lambda_{x}I_x + \mu O'_R$
is a positive operator.
Let $ S_y\in \Evd {y} $ be an arbitrary deterministic event of type $y$.
By the induction hypothesis we have that  $ S_y= \lambda_y I_y
+ Y$, where $Y \in \set{\Delta}_y$. By positivity of $R'$ one has
$0 \leq \Tr_y[(S_y \otimes I_z)  R' ]$. By direct computation, one can show that $\Tr_y[(S_y \otimes I_z)  R' ]= \lambda_z I_z + Z$
for some $Z \in  \set{\Delta}_z  $. By the induction hypothesis we have
$\Tr_y[(S_y \otimes I_z)  R' ]\in \Evd {z}$.
This proves the following inclusion:
$\set{\Delta}_{x} \supseteq [\set{Herm}(\hilb{H}_y) \otimes {\set{\Delta}_{z}}]
    \oplus
[\overline{\set{\Delta}}_{y}   \otimes
{\set{\Delta}}^\perp_{z} ]$. Thus,
$\set{\Delta}_{x} =[\set{Herm}(\hilb{H}_y) \otimes {\set{\Delta}_{z}}]
    \oplus
[\overline{\set{\Delta}}_{y}   \otimes 
{\set{\Delta}}^\perp_{z} ]$.\qed

\begin{corollary}\label{cor:equivreformulated}
  Let $x$ and $y$ be two types.
  Then we have
\begin{align}
  x \equiv y \iff
  \lambda_x = \lambda_y \land \Delta_x = \Delta_y 
      \label{eq:equivreformlation}
\end{align}
\end{corollary}

\begin{corollary}\label{lem:identityop}
  Let $x$ be a type and let
  $A_i$ denote the elementary types occurring in the definition of $x$.
  Let $I_x$ be the identity operator in
  $\mathcal{L}(\hilb{H}_x)$ and let $\lambda_x$ be defined as in Equation~\eqref{eq:lambdarecursive}. Then we have
\begin{align}
  &\lambda_x I_x \in \Evd {x}, \qquad
    \lambda_x = \prod_{A_i \in x} d_{A_i}^{-K_x(A_i)} \\
&  K_x(A_i) :=  \# [``\to" ] + \# [``(" ]  \pmod2
      \label{eq:identityisadm}
\end{align}
$\# [``\to" ]$ and $\# [``(" ]$ denotes the number of arrows $\to$
and open round brackets $($ to the right of $A_i$ in the expression of
$x$, respectively.
\end{corollary}
\Proof
The only non-trivial claim is that
$\lambda_x = \prod_i d_i^{-K(i)} $.
Let us prove this statement by induction.
The thesis is true for elementary types.
We now suppose that the thesis holds for any $y\preceq x$, and
we consider a non elementary type $x = y \to z$.

Let $A$ be any elementary type occuring in the expression of $y$.
We now show that $K_x(A)$ is $K_y(A) + 1 \pmod 2$.
First we observe that the expression of type $y$ must occurs in
$y \to z$ with the outermost parenthesis. By definition
\ref{def:quantum-types}, we have that 
any expression of a type, with the outermost parenthesis, contains as
many ``$\to$'' as ``$($''. 
Then we consider an elementary type $A$ which occurs in the
definition of $y$. The same type will occur in the expression of $y
\to z$. It is easy to realize that the number of ``$\to$'' and ``$($'' that follow
$A$ in the expression of $y \to z$ is changed by an odd number.
Indeed, we now have all the ``$\to$'' and ``$($'' that appear in
$y$ plus one more $\to$ which is the $\to$ that stays between $y$ and
$z$.

Let us now consider
 an elementary type $B$ which occurs in the
 expression of $z$. The same $B$  appears in the expression of $y
 \to z$ and  the number  ``$\to$'' and ``$($'' to its right is
 unchanged and therefore $K_x(B)=K_z(B)$.
 Then we have
 \begin{align*}
 \prod_{A_i \in x} d_{A_i}^{-K_x(A_i)} &=  \prod_{A_i \in y}
   d_{A_i}^{-[K_y(A_i) +1 \!\! \! \! \! \!  \pmod 2]}    \prod_{A_i \in z}
   d_{A_i}^{-K_z(A_i)} =\\
&=  \left( d_y  \prod_{A_i \in y}
   d_{A_i}^{-K_y(A_i) }  \right)^{-1}  \prod_{A_i \in z}
   d_{A_i}^{-K_z(A_i)} 
 \end{align*}
which proves that recurrence relation of
Equation~\eqref{eq:lambdarecursive}
is satisfied. \qed

\begin{corollary}
\label{cor:char-high-order-set}
  For any type $x$,  we have 
  \begin{align}
    \label{eq:deltaisthedirectsum}
    \set{\Delta}_x = \bigoplus_{\bvec{b} \in D_x}  \set{L}_{\bvec{b}}
  \end{align}
for some set $D_x$ of string.
\end{corollary}
\Proof
The thesis is true for elementary types ($D_E = 0$).
Let us suppose that $\set{\Delta}_x$ and $\set{\Delta}_y$ are
  the direct sum of $\set{L}_{\bvec{b}}$ spaces
for two  types $x$ and $y$.
Then, by Equation \eqref{eq:eledetchar2},
also
$\set{\Delta}_{x\to y }$ is
the direct sum of $\set{L}_{\bvec{b}}$ spaces. \qed
  Notice that the expression on the right hand side of Eq.~\eqref{eq:deltaisthedirectsum} can involve different choices of $D_x$ depending on the number of trivial systems $I$ that are explicitly considered in the expansion of $\hilb H_x$. However, the space $\set \Delta_x$ on the right hand side is uniquely defined, independently of the choice of $D_x$. In particular, there is one preferred choice for $D_x$ which is the one obtained after reducing the strings $D_x$ to their normal form $(D_x)'_N$ as in Eq.~\eqref{eq:normform}. It is easy to realise that the set $(D_x)'_N$ 
depends only on the natural type structure of the type $x$.
If two types $x$ and $x'$ have the same natural type structure, then we have 
\begin{align}
  \label{eq:defofDforstructure}
\mathrm{x}:= [x] = [x'] \implies D_{\rm{x}} := (D_x)'_N =  (D_{x'})'_N  .  
\end{align}
Moreover, it is possible to generalise
Proposition~\ref{prop:char-high-order}
to type structures.
\begin{corollary}
\label{cor:char-high-order-setrecursive}
Let ${\rm{x}}$ be a type structure and
let $D_{\rm{x}}$  be the set of strings defined according to Equation~\eqref{eq:defofDforstructure}.
Then $D_{\rm{x}}$ is such that:
\begin{align}
  \label{eq:charrecursiveset}
\begin{aligned}
 &D_{I} = \emptyset \quad  D^{\perp}_{I} = \{{\boldsymbol{\varepsilon}}\}, \quad  D_{*} = \{0\}, \\
  &D_{{\rm{x}} \to {\rm{y}}}= W_{{\rm{x}}}D_{\rm{y}}
  \cup
  \overline{D}_{\rm{x}} D^{\perp}_{\rm{y}}, 
  \end{aligned}
\end{align}
where the sets $W,T$ have been defined in
Equation~\eqref{eq:tracelessspace}, the sets $D$ have been defined in
Equation~\eqref{eq:defofDforstructure} and juxtaposition of sets and
strings has been defined in Equation~\eqref{eq:concatset}.
\end{corollary}

 {The results that we presented in this section are the basic technical 
tools in the study of higher-order quantum maps.
In particular, Proposition \ref{prop:char-high-order} unfolds
the characterization of admissible events given in
Theorems~\ref{prop:positiveconeadmis} and ~\ref{prop:deteventmadeeasy},
and provides an explicit constructive formula.
In the next subsections we will apply this result to prove some
equivalence between types (and type structures).}

  \subsection{Functionals}\label{sec:functionals}

  In this section we study the types of the kind
$x \to I$ (we remind that $I$ denotes the type of the trivial
elementary system).
Events of type $x \to I$
are linear functionals on events of of type $x$.
It is convenient to introduce the shorthand notation
\begin{align}
  \label{eq:functionalnotation}
  \overline{x} :=  x \to I.
\end{align}
By virtue of Proposition
\ref{prop:char-high-order}
we have the following lemma.
\begin{lemma}
\label{lem:tracelessfunctionals}
  Let $x$ be a type and let $\set{\Delta}_x$ and $\lambda_{{x}}$
  be defined as in Proposition
  \ref{prop:char-high-order}.
  Then $\set{\Delta}_{\overline{x}} =\overline{\set{\Delta}}_{{x}} $
  and $\lambda_{\overline{x}} = \frac{1}{\lambda_{{x}} d_x }$.
\end{lemma}
\Proof
For the trivial system $I$, we have
$\lambda_I =1$
and 
$\set{\Delta}_{I} = 0$.
Then, from Equation
\eqref{eq:eledetchar2}
we immediately have
\begin{align*}
  \set{\Delta}_{x \to I} = [
  \overline{\set{\Delta}}_{x} \otimes \set{Herm}(\mathbb{C}) ] = \overline{\set{\Delta}}_{x}
\end{align*}
and
$\lambda_{\overline{x}} = \frac{1}{\lambda_{{x}} d_x }$. \qed
We can now easily prove the following identity.
\begin{proposition}
\label{prp:involution}
  Let $x$ be a type. Then $\overline{\overline{x}} \equiv x$.
\end{proposition}
\Proof
By definition \ref{def:equivtyp}, $\overline{\overline x}\equiv x$ iff $\Evd x =
\Evd {\overline{\overline{x}}}$. Now,
\begin{align*}
  R \in \Evd x \iff R \geq 0, \quad R = \lambda_xI  + X, \quad
X \in \set{\Delta}_x.
\end{align*}
Using Equation \eqref{eq:eledetchar2}
we have
\begin{align}
  \label{eq:auxiliaryequation}
  \lambda_{\overline{\overline x}}=\lambda_{(x \to I)\to I} = \frac{1}{d_{x\to I} \lambda_{x \to I}}=
   \frac{1}{d_{x}   \frac{1}{d_x \lambda_x}} = \lambda_x.
\end{align}
Then we have
\begin{align*}
  R \in \Evd {\overline{\overline{x}} } &\iff
  R \geq 0, \quad  R = \lambda_{\overline{\overline{x}} }I_{\overline{\overline x}}   +
  X, \quad  X
  \in  \set{\Delta}_{\overline{\overline{x}} } \\
                                               &\iff
                                                 R \geq 0, \quad
  R = \lambda_{x}  I_x  +
  X,  \quad  X
  \in  \overline{\overline{\set{\Delta}}}_{{x}} \\
                                               &\iff
                                                 R \geq 0, \quad
  R = \lambda_{x}  I_x  +
  X,  \quad  X
  \in \set{\Delta}_{{x}} \\
  & \iff R \in \Evd x,
\end{align*}
where we used Equation \eqref{eq:auxiliaryequation} and
Lemma \ref{lem:tracelessfunctionals}
in the second line. \qed
Let us clarify the previous discussion with some examples.
We know that for an elementary type $A$ the set $\Evd A$
is the set of positive operators on $\hilb{H}_A$ with unit trace,
(i.e. $\set{\Delta}_A = \set{Traceless}(\hilb{H}_A)$ and $\lambda_A =
d_A^{-1}$),  while for the trivial elementary type $I$ we have
$\Evd I = 1$ (i.e. $\set{\Delta}_I = \set{Traceless}(\mathbb{C})=0$ and $\lambda_I =
d_I^{-1}=1$). 
By applying Equation \eqref{eq:eledetchar2}
we have $\lambda_{A \to I} = 1$
and $\set{\Delta}_{A \to I} = 0$. Indeed, 
$\set{\Delta}_{A \to I} =[ \overline{\set{\Delta}}_{A} \otimes \set{Herm}(\mathbb{C}) ]
    \oplus
[ ( \set{L}_{\bvec{1}}  \oplus
\set{\Delta}_{A} )  \otimes  {\set{\Delta}_{I}}    ] =
[ \overline{\set{Traceless}(\hilb{H}_A)}  \otimes   \set{Herm}(\mathbb{C})    ]
    \oplus
[   ( \set{L}_{\bvec{1}}  \oplus
\set{Traceless}(\hilb{H}_A) )  \otimes {\set{\Delta}_{I}}  ] = 0 $,
since
$\set{\Delta}_{I} = \overline{\set{Traceless}(\hilb{H}_A)}=0$.
Then we have
$\Evd {A \to I} = I_A$, i.e.
the set of deterministic events of type $A \to I$
has only one element, the identity operator on
$\hilb{H}_A$.
The set of probabilistic events of type 
$A \to I$ is the set of positive operators bounded by $I$.
We recover then the usual notion of effect (element of a POVM).
The equivalence $\overline{\overline{A}} \equiv A$
tells us that a quantum state can be equivalently
interpreted as the Choi operator of a map that sends a deterministic
measurement (which is uniquely represented by the identity operator)
to the number $1$.
It seems we have gone quite a long and devious way to 
prove an obviuos fact. However, as we will see, when considering
more complex types, the equivalence between types can be far from obvious.

\subsection{Tensor product of types}\label{sec:tensor-product-types}
In this section we introduce the following composition law for types:
\begin{align}
  \label{eq:tensortypes}
  x \otimes y := \overline{x \to \overline{y}}.
\end{align}
This operation can be thought of as the generalization of parallel
composition of elementary types to the whole hierarchy.
\begin{lemma}[Characterization of tensor product of types]
  \label{lmm:tensor-product-type}
  Let $x$ and $y$ be two types
  and let $\set{\Delta}_x,\lambda_{{x}}, \set{\Delta}_y,\lambda_{{y}}$
  be defined as in Proposition
  \ref{prop:char-high-order}.
  We have:
  \begin{align}
    \label{eq:tensorchara}
    \begin{aligned}
    \set{\Delta}_{x \otimes y} &=
    (\set{L}_{\bvec{e}} \otimes \set{\Delta}_{x})  \oplus
(\set{\Delta}_{y} \otimes \set{\Delta}_{x})  \oplus
(\set{\Delta}_{y}\otimes \set{L}_{\bvec{e}})  \\
\lambda_{x \otimes y} &= \lambda_{x}\lambda_{y}
    \end{aligned}
  \end{align}
\end{lemma}
\Proof The thesis can be easily proved by recursively applying
Equation \eqref{eq:eledetchar2}. \qed
\begin{proposition}[Properties of the tensor product of types]
  The following equivalences hold:
  \begin{align}
    \label{eq:equivalencestensors}
 &   A \otimes B \equiv AB, \quad \forall \; A, B \in \ETy\\
  &  x \otimes y \equiv y \otimes x, \quad \forall \; x, y \in \set{Types} \\
    &(x \otimes y) \otimes z \equiv x \otimes (y \otimes z) , \quad
      \forall \; x, y ,z \in \set{Types}
      \label{eq:associativitytensortype}
  \end{align}
\end{proposition}
\Proof
By recursively applying
Equation \eqref{eq:tensorchara}
one has
$\lambda_{ A \otimes B} = \lambda_{AB}$,
$\lambda_{ x \otimes y} = \lambda_{y \otimes x}$,
$\lambda_{ (x \otimes y) \otimes z} = \lambda_{x \otimes (y \otimes
  z)}$,
$\set{\Delta}_{ A \otimes B} = \set{\Delta}_{AB}$,
$\set{\Delta}_{ x \otimes y} = \set{\Delta}_{y \otimes x}$ and
$\set{\Delta}_{ (x \otimes y) \otimes z} = \set{\Delta}_{x \otimes (y \otimes
  z)}$.
Since the cone of positive operators
depends only on the elementary systems
occurring in the definition of a type,
we have
$\set{P}_{ A \otimes B} = \set{P}_{AB}$,
$\set{P}_{ x \otimes y} = \set{P}_{y \otimes x}$ and
$\set{P}_{ (x \otimes y) \otimes z} = \set{P}_{x \otimes (y \otimes
  z)}$
and the thesis follows. \qed

We have seen that the tensor product of elementary types recovers the
familiar notion of tensor product of quantum systems.
However, when non trivial types are involved,
the interpretation of the tensor product between two types is
more subtle.
Let us clarify this feature with an example.
Let us consider  the types $A\to B$
and $C \to D$.
The deterministic events of type
$A\to B$
and $C \to D$
are quantum channels from system
$A$ to system $ B$
and
quantum channels from system
$C$ to system $ D$,
respectively.
Then we have
\begin{align*}
  \begin{aligned}
&R \in \Evd {A \to B} \iff
  R = \frac{1}{d_B}I + X  \\ 
 & R \in \set{P}(\hilb{H}_A\otimes \hilb{H}_B), \\
  &X \in \set{L}_T
  \otimes \set L_W,
  \end{aligned}
\end{align*}
An analogous equation holds for $C \to D$.
Let us now consider the type
$(A \to B) \otimes (C \to D) $.
From Equation \eqref{eq:tensorchara}
we have that $ R \in \Evd {(A \to B) \otimes (C \to D)}$ iff
\begin{align}
  \label{eq:softcomb}
   R =& \frac{1}{d_D d_B}I + X, \quad R \geq 0,\nonumber \\ 
    X \in\; &
    [\set L_W
    \otimes
 \set{L}_T
    \otimes
    {\set{L}_{\bvec{e}}}
    \otimes
    {\set{L}_{\bvec{e}}}]
    \oplus\\
    &
    [{\set{L}_{\bvec{e}}}
    \otimes
    {\set{L}_{\bvec{e}}}  
    \otimes
    \set{L}_W
    \otimes
  \set{L}_T]
    \oplus\nonumber\\
    &
   [\set L_W
 \otimes
     \set{L}_T
    \otimes \set L_W\otimes
    \set{L}_T].\nonumber
\end{align}
Operators that obey Equation \eqref{eq:softcomb}
are Choi operators of \emph{non-signalling channels},
\begin{align}
  \label{eq:nosignallingdiagram}
  \mathcal{R}:\mathcal{L}(\hilb{H}_A \otimes \hilb{H}_C) \to
  \mathcal{L}(\hilb{H}_B \otimes \hilb{H}_D),\qquad 
  \begin{aligned}
\Qcircuit@C=0.5em @R=1em {
    \ustick{\scriptstyle{A}}&\multigate{1}{{R}}&\ustick{\scriptstyle{B}}\qw&\\
   \ustick{\scriptstyle{C}}  &\ghost{{R}}&\ustick{\scriptstyle{D}} \qw&}
  \end{aligned},
\end{align}
which send
quantum states of the bipartite system $AC$
to quantum states of the bipartite  system $BD$,
such that the output $B$ does not depend on the input
$C$
and
the output $D$ does not depend on the input
$A$.
Non-signalling channels of this kind have two possible
realisations as memory
channels
as follows \footnote{further details about the realization of
  no-signalling bipartite channels can be found in Ref.~\cite{PhysRevLett.106.010501}.}:
\begin{align*}
   \begin{aligned}
\Qcircuit@C=0.5em @R=1em {
    \ustick{\scriptstyle{A}}&\multigate{1}{{R}}&\ustick{\scriptstyle{B}}\qw&\\
   \ustick{\scriptstyle{C}}  &\ghost{{R}}&\ustick{\scriptstyle{D}} \qw&}
\end{aligned}
 =
  \begin{aligned}
\Qcircuit @C=0.5em @R=1em {
  \ustick{\scriptstyle{A}}&
  \multigate{1}{{R}_1}&
  \ustick{\scriptstyle{B}}\qw&
  &
  \ustick{\scriptstyle{C}}&
  \multigate{1}{{R}_2}&
  \ustick{\scriptstyle{D}}\qw&
  \\
  &
  \pureghost{{R}_1}&
  \qw&
  \qw&
  \qw&
  \ghost{{R}_2}&&}
\end{aligned}
                                  =
                                   \begin{aligned}
\Qcircuit @C=0.5em @R=1em {
  \ustick{\scriptstyle{C}}&
  \multigate{1}{\tilde{R}_1}&
  \ustick{\scriptstyle{D}}\qw&
  &
  \ustick{\scriptstyle{A}}&
  \multigate{1}{\tilde{R}_2}&
  \ustick{\scriptstyle{B}}\qw&
  \\
  &
  \pureghost{\tilde{R}_1}&
  \qw&
  \qw&
  \qw&
  \ghost{\tilde{R}_2}&&}
  \end{aligned}.
\end{align*}
The previous equation means that
for any non signalling channel
$\mathcal{R}: \mathcal{L}(\hilb{H}_A \otimes \hilb{H}_C) \to
  \mathcal{L}(\hilb{H}_B \otimes \hilb{H}_D)$ there exist
four channels
$\mathcal{R}_1,\tilde{\mathcal{R}}_2 : \mathcal{L}(\hilb{H}_A) \to
\mathcal{L}(\hilb{H}_B)$ and
$\mathcal{R}_2,\tilde{\mathcal{R}}_1: \mathcal{L}(\hilb{H}_C) \to
\mathcal{L}(\hilb{H}_D)$ such that
$\mathcal{R}$ can be realized as either the concatenation of
$\mathcal{R}_1$ and $\mathcal{R}_2$
or the concatenation
of 
$\tilde{\mathcal{R}}_1$ and $\tilde{\mathcal{R}}_2$.
Given two
channels $\mathcal{R} : \mathcal{L}(\hilb{H}_A) \to
\mathcal{L}(\hilb{H}_B)$ and
$\mathcal{S} : \mathcal{L}(\hilb{H}_C) \to
\mathcal{L}(\hilb{H}_D)$,
their tensor product
$\mathcal{R} \otimes
\mathcal{S} $
is a non-signalling channel.
Also the convex combination
$p\mathcal{R} \otimes
\mathcal{S} + (1-p )
\mathcal{R}' \otimes
\mathcal{S}'$
of tensor product of channels
is a non-signalling channel.
However not every
non-signalling channel
is a convex combination
of tensor products of channels.
In the language of  {higher-order quantum theory}, this means that the
following strict inclusion holds:
\begin{align*}
  \begin{aligned}
      \Evd {x \otimes y}
&= \set{P}(\hilb{H}_x\otimes \hilb{H}_y)
\cap
  \set{Aff} \{ \Evd {x} \otimes
  \Evd {y} \}  
  \supset \\
&\supset  \set{Conv} \{ \Evd {x} \otimes
  \Evd {y} \}  
  \end{aligned}
\end{align*}
where
 $\set{Aff} \{ S \}$
denotes the affine hull of the set $S$
and
 $\set{Conv} \{ S \}$
denotes the convex hull of the set $S$.

We conclude this subsection by proving the
\emph{uncurrying} identity for higher-order quantum maps
\begin{proposition}[Quantum uncurrying]
  \label{prp:curryingident}
  For any types $x$ , $y$ and $z$ we have the equivalence
  \begin{align}
    \label{eq:currying}
   x  \to (y \to z)\equiv (x \otimes y) \to z
  \end{align}
\end{proposition}
\Proof
The equivalence \eqref{eq:currying}
is consequence of the associativity of the tensor product of types 
Indeed, from Equation \eqref{eq:associativitytensortype} and
Proposition \ref{prp:involution}
we have
\begin{align*}
&\overline{(x \otimes y) \to \overline{z}} \equiv
\overline{x \to \overline{(y \otimes z)} }   \\
&\iff
(x \otimes y) \to \overline{z}
\equiv
x \to (\overline{y \otimes z}) \\
&\iff
 (x \otimes y) \to \overline{z}
\equiv
x \to ({y \to \overline{z}}). 
\end{align*}
By substituting $\overline{z}$
with $z$ we have he thesis. \qed

\subsection{Generalized comb}

In this subsection, we study the following family of
sub-hierarchies:
\begin{definition}[n-comb with base $\mathrm x$]\label{def:generalized-comb}
  Let $\mathrm x$ be a type structure. The type structure
  $\mathbf{n}_{\mathrm x}$ of \emph{n-combs with base 
    $\mathrm x$} is defined recursively as follows:
  \begin{itemize}
  \item $\mathbf{1}_{\mathrm x} =\mathrm{x} $,
    \item $\mathbf{n}_{\mathrm x} = (\mathbf{n-1})_{\mathrm x}\to \mathrm x $.
    \end{itemize}
    The type structure $\mathrm x$ is called the \emph{base} of
    the type structure $\mathbf{n}_{\mathrm x} $.
    We denote with $\mathbf{n}_{x}$ a generic type such that
    $\mathbf{n}_{\mathrm x} $ is its natural type structure, i.e.
    $[\mathbf{n}_{x}] = \mathbf{n}_{\mathrm x} $.
  \end{definition}  
  According to Definition \ref{def:generalized-comb} a type
  $\bvec{n}_x$ has the following expression:
  \begin{align}
    \label{eq:combexpression}
    \begin{split}
    &\bvec{n}_x  = ((\dots((x_1 \to x_2)\to x_3) \dots )\to x_{n-1})\to
    x_n, \\
    &[x_i] = \mathrm{x} \quad \forall i =1 , \dots ,n.
    \end{split}
  \end{align}
For   example,
  $\mathbf{4}_\mathrm x= ((\mathrm x\to \mathrm x)\to \mathrm x)\to
  \mathrm x$ and $\mathbf{4}_x = ((x_1 \to x_2)\to x_3)\to
  x_4$.  As it is known, the case in which
  $\mathrm x = *\to*$ gives rise to the \emph{comb}
  hierarchy which is extensively studied in the
  literature\cite{PhysRevLett.101.060401,PhysRevA.80.022339,bisio2016quantum,PhysRevA.77.062112,Jencova:2012aa}.


As it will be soon clear, the language of type structures, which
  was unnecessary in subsections~\ref{sec:functionals} and
  \ref{sec:tensor-product-types}, simplifies the study of the quantum
  types introduced by Definition~\ref{def:generalized-comb}. 
 Our first result is a characterisation theorem for n-combs of base 
  $\mathrm x$. 

  \begin{proposition}[Characterisation of generalized n-combs]
    \label{prp:charageneralized-comb}
Let $\mathbf{n}_{\mathrm x} $ be a type structure defined as in
Definition~\ref{def:generalized-comb}.
Then we have
\begin{align}
       \label{eq:charcombsubspace}
  &  D_n =
     \begin{cases}
       \begin{aligned}
       \bigcup_{l=1}^{\tfrac{n+1}{2}} & W^{n-2l+1} D {D^{\perp}}^{2l-2}
       \\
       &\cup \   \bigcup_{l=1}^{\tfrac{n-1}{2}} \, \bvec{e}^{2l-1} \overline{D}
         {D^{\perp}}^{n-2l}
       \end{aligned}
       &n \mbox{ odd}\\
       \\
        \begin{aligned}
       \bigcup_{l=1}^{\tfrac{n}{2}} & \left( W^{n-2l-1} D {D^{\perp}}^{2l-2}
        \right. \\
       & \left.  \cup \, \bvec{e}^{2l-2} \overline{D}
         {D^{\perp}}^{n-2l+1} \right)
       \end{aligned}
      &n \mbox{ even}
      \end{cases}
\end{align}
where the sets $D_n$ are defined according to
    Equation~\eqref{eq:defofDforstructure} for the type structure
    $\bvec{n}_{\rm{x}}$ with $D:=D_1$, and  $W := W_{\rm{x}}$,
    $\bvec{e}:= \bvec{e}_{\rm{x}}$ are
    defined according to Equation \eqref{eq:tracelessspace}.
    Moreover, for any type $\bvec{n}_x $, 
    we have
\begin{align}
  \label{eq:charcomblambda}  
  \lambda_n =
   \begin{cases}
     \lambda_{x_n}\prod_{i=1}^{\tfrac{n-1}{2}} \left[\lambda_{x_{2i-1}}
     \left(\lambda_{x_{2i}} d_{x_{2i}} \right)^{-1}\right] &n \mbox{ odd}\\
        \\
\prod_{i=1}^{\tfrac{n}{2}} \left[\lambda_{x_{2i}}
     \left(\lambda_{x_{2i-1}} d_{x_{2i-1}} \right)^{-1}\right] 
        &n \mbox{ even}.
      \end{cases}
\end{align}
       where $\lambda_n$ is defined as in
       Proposition~\ref{prop:char-high-order}
       and $x_i $ are defined as in Equation~\eqref{eq:combexpression}.
  \end{proposition}
  \Proof
Let us begin with the proof of Equation
\eqref{eq:charcombsubspace}.
The thesis hold for $\bvec{1}_{\rm{x}}$.
Let us then suppose that the thesis holds for any $m < n+2$ and $m$ even.
By applying corollary~\ref{cor:char-high-order-setrecursive}
twice, we have 
\begin{align*}
  D_{n+2}
  &= W_{n-1}D_1 \cup \overline{D}_{n-1}D^{\perp}_1 =\\
  & =  W_{n-1}D_1 \cup \bvec{e}_{n-2} D_1 D^{\perp}_1 \cup D_{n-2} D^{\perp}_1 D^{\perp}_1
\end{align*}
which, thanks to the induction hypothesis, proves the thesis for $n$
even.
The proof for $n$ odd is analogous.

We now focus on Equation
  \eqref{eq:charcomblambda}.
  Since $\bvec{1}_{x} ={x_1}$ the thesis clearly holds.
Let us fix an arbitrary odd $n$ and
  let us suppose that the thesis hold for any $m < n$.
  Since $\bvec{n}_{x}=(\bvec{n-1})_{x} \to {x_n}$,
  by combining Equation \eqref{eq:charcomblambda}
  and the induction hypothesis, we have
  \begin{align*}
    \lambda_{n} &= \frac{\lambda_{x_n}}{d_{n-1}\lambda_{n-1}} \\
    &=  \frac{\lambda_{x_n}}{\prod_{i=1}^{\tfrac{n-1}{2}} \left[
                    d_{x_{2i}}d_{x_{2i-1}} \lambda_{x_{2i}}
    \left(\lambda_{x_{2i-1}} d_{x_{2i-1}} \right)^{-1}\right] } =\\
&= \frac{\lambda_{x_n}}{\prod_{i=1}^{\tfrac{n-1}{2}} \left[
                    d_{x_{2i}}\lambda_{x_{2i}}
    \left(\lambda_{x_{2i-1}} \right)^{-1}\right] }     =\\
  &= \lambda_{x_n}\prod_{i=1}^{\tfrac{n-1}{2}} \left[
           \lambda_{x_{2i-1}}         (\lambda_{x_{2i}} d_{x_{2i}} )^{-1}
    \right],     
  \end{align*}
  which proves  the thesis for odd $n$.
The $n$ even case can be proved by a similar calculation.
\qed


In order to clarify the discussion, it is convenient to analyze some
examples in detail.
Let us start with the case in which the base  $\rm{x} $
is the elementary structure, i.e $\rm{x}  = *$,
and 
\begin{align}
  \bvec{n}_E   = (\dots((E_1 \to E_2)\to E_3) \dots )\to
    E_n, 
\end{align}
Then we have
\begin{align}
  &D = \{0\}, \quad \overline{D} = \emptyset , \quad D^{\perp} = \{1\} \nonumber \\
  &\bvec{b} \in D_n \iff
  \begin{aligned}
    &\bvec{b}  \mbox{ starts from the right} \\
    &\mbox{with an
  even number of 1s},   
\end{aligned}
  \label{eq:normacomusual} \\
&  \lambda_n =
   \begin{cases}
     d^{-1}_{E_n}\prod_{i=1}^{\tfrac{n-1}{2}}d^{-1}_{E_{2i-1}}  &n \mbox{ odd}\\
        \\
\prod_{i=1}^{\tfrac{n}{2}} d^{-1}_{E_{2i}}
        &n \mbox{ even}.
      \end{cases}
\end{align}

Then, let us analyze the comb hierarchy, i.e. the case
$\mathrm{x} = * \to *$,
\begin{align}
  \label{eq:combtypeexpression}
  \begin{split}
  \bvec{n}_{A \to B}   
  =(\dots (A_1 \to B_1)\to) \dots )\to
    (A_n \to B_n).   
  \end{split}
\end{align}
We have
\begin{align*}
&W = \{  00,01,10,11 \}\\
   &D = \{ 10,00 \}, \quad \overline{D} =\{  01 \}, \quad D^{\perp} =
    \{11 , 01 \}  
\end{align*}
From Proposition~\ref{prp:charageneralized-comb} we have that
$D_n$ has the following structure 
\begin{align*}
  \begin{split}
  D_n &= W_{n-1}D \cup W_{n-3}D {D^{\perp  }}^2\cup \dots \cup
  D{D^{\perp }}^{n-1} \cup \\
  & \bvec{e}
  \overline{D}{D^\perp}^{n-2} \cup  \bvec{e}^{3}
  \overline{D}{D^\perp}^{n-4} \dots \cup
  \bvec{e}^{n-2} \overline{D}{D^\perp} \; \mbox{(}n  \mbox{ odd)},  
    \end{split} \\
    \begin{split}
      D_n &= W_{n-1}D \cup W_{n-3}D {D^{\perp  }}^2\cup \dots \cup
  WD{D^{\perp }}^{n-2} \cup \\
  &\overline{D}{D^\perp}^{n-1} \cup \bvec{e}^2
  \overline{D}{D^\perp}^{n-3} \cup \dots \cup
  \bvec{e}^{n-2} \overline{D}{D^\perp} \; \mbox{(}n  \mbox{ even)},  
    \end{split}
\end{align*}
for example, for
$\bvec{3}_{A \to B}$ we have:
$    D_3 $ $= WWD $ $\cup$ $ WDD^{\perp}D^{\perp}
    $ $\cup $ $\bvec{e} DD^{\perp}  $.
We see that type structure
of $\bvec{n}_{E \to E}$
induces a decomposition of the binary string $\bvec{b}$
into $n$ binary strings of two digit, i.e.
\begin{align*}
  \begin{aligned}
  &\begin{aligned}
 \bvec{b} =& \bvec{w}_1 \bvec{w}_2 \dots \bvec{w}_n =\\
 =&w^A_1\, w^B_1 \,\,\, w^A_2 \,w^B_2 \,\,\dots\,\, w^A_n\, w^B_n
  \end{aligned}\\
 &  w^{E}_i = 0,1, \quad i=1,\dots n, \quad E=A,B.
  \end{aligned}
\end{align*}
Let us then consider the   following permuted string
\begin{align*}
  \tilde{\bvec{b}} := w^A_n\, \dots \,  w^A_2 \, w^A_1 \,  w^B_1\,
  w^B_2 \, \dots \, w^B_n.
\end{align*}
Thanks to Equation~\eqref{eq:normacomusual}, we have
\begin{align}
  \label{eq:equivalence}
  \begin{aligned}
  \bvec{b} \in D^{A\to B}_n & \iff    \begin{aligned}
    & \tilde{\bvec{b}}   \mbox{ starts from the right} \\
    &\mbox{with an
      even number of 1s},
    \end{aligned} \\ &\iff
    \tilde{\bvec{b}} \in D^{E}_{2n}   
  \end{aligned}
\end{align}
where the superscript to the sets $D_n$
reminds us that we are considering two different comb hierarchies.
We can therefore prove the following equivalence between types.
\begin{proposition}[Equivalence between $\bvec{n}_{A \to B}$ and
  $\bvec{2n}_{E}$]
  Let $A_i$ , $B_i$ , $1\leq i \leq n$ be elementary types. 
  Then following equivalence holds:
  \begin{align}
    & ( \cdots ((A_1 \to B_1 )\to (A_2\to B_2))\cdots )\to (A_n \to B_n)
     \equiv \nonumber \\
  &\equiv ( \cdots (A_n \to
    A_{n-1})\cdots \to A_1) \to B_1 ) \cdots) \to B_n
    \label{eq:equivacomb}
\end{align}
\end{proposition}
\Proof The identity $\set{\Delta}_n^{A \to B}$ and
$\set{\Delta}_{2n}^{E}$ follows from Equation \eqref{eq:equivalence},
which holds unchanged also in the more general case in which the
elementary types have different dimensions.  Then, with the help of
Equation~\eqref{eq:charcomblambda} one can verify that
$\lambda_n^{A\to B}= \lambda_{n2}^E $. \qed 
The proof of Equation~\eqref{eq:equivalence}, which leads to the non trivial type
equivalence~\eqref{eq:equivacomb}, is an example highlighting the relevance of the formalism introduced in
Subsection~\ref{sec:setl_bvecb--spaces}.

We now further investigate the comb hierarchy of types $\bvec{n}_{A
  \to B}$. From this point to the end of this subsection,
the subscripts $n$ or $m$ or $p$ will refer to the comb hierarchy,
namely the types $\bvec{n}_{A\to B}$.

From the type equilavence of Equation \eqref{eq:equivacomb}
and from Equation \eqref{eq:normacomusual},
we recover the usual normalization condition for comb:
\begin{align*}
  &R^{(n)} \in \Evd {\bvec{n}_{A \to B}} \!  \! \iff
  \begin{cases}
    R^{(n)}\geq 0 \\
    \Tr_{{2k}}[R^{(k)}] = I_{{2n-1}} \otimes R^{(k-1)}\\
R^{(0)}= 1 ,\quad  k = 1, \dots, n ,
\end{cases}\\
&  E_i =
  \begin{cases}
    A_{n-i+1} &   1 \leq i \leq n\\
    B_{i-n}  &  n+1 \leq i \leq  2n,   
  \end{cases}
\end{align*}
where $\Tr_{{i}}$ and $I$ denote the partial trace
and the identity operator on the Hilbert space of
the system
$E_i$. As it is well known,
$n$-comb can be realized as
causally order \emph{quantum network} with
$n$ vertices (i.e. a sequence of channels with memory).
For example we have
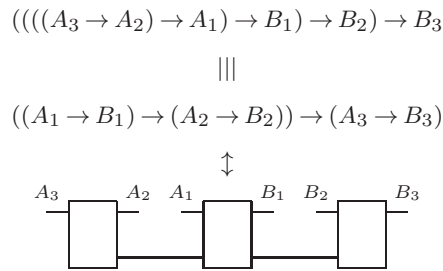
\begin{gather*}
    ((((A_3 \to A_2) \to A_1)\to B_1)\to B_2)\to B_3  
 \\[1pt]
|||    
  \\[1pt]
    ((A_1 \to B_1 )\to (A_2\to B_2))\to (A_3\to B_3)  
    \\[1pt]
  \updownarrow    
  \\[1pt]
  \Qcircuit @C=0.9em @R=1em {
  \ustick{\scriptstyle{A_3}}&
                            \multigate{1}{\quad}&
                                                          \ustick{\scriptstyle{A_2}}\qw&
&
  \ustick{\scriptstyle{A_1}}&
                            \multigate{1}{\quad}&
                                                          \ustick{\scriptstyle{B_1}}\qw&
                                                                                                         &
                                                                                                          \ustick{\scriptstyle{B_2}}&
                                                                                                                                       \multigate{1}{\quad}&
                                                                                                                                                                     \ustick{\scriptstyle{B_3}}\qw\\
                          &
                            \pureghost{\quad}&
                                                       \qw&
                                                            \qw&
                                                                 \qw&
                                                                      \ghost{\quad}&
                                                                                             \qw&
                                                                                                  \qw&
                                                                                                                           \qw          &
                                                                                                                                       \ghost{\quad}&}        
\end{gather*}

Thanks to Equation \eqref{eq:equivalence}
it easy to prove that, for $p = n+m  $,
\begin{align}
  \label{eq:combsum}
  \begin{aligned}
  D_p =&
  {\bvec{e}}_{\, n}  D_m
  \cup D_n  \bvec{e}_{\, m} \cup\\
 & \cup
  D_n D_m 
\cup
  \overline{D}_n  \otimes D_m.
    \end{aligned}
\end{align}
By combining
Equation~\eqref{eq:tensorchara}
and Equation~\eqref{eq:combsum}
we obtain the characterization of the type
$\bvec{n} \otimes \bvec{m}$, i.e.
\begin{align}
  \label{eq:combtensor2}
  \begin{aligned}
  \set{\Delta}_{m \otimes n} =&
  {\set{L}_{\bvec{e}}}_{\, n} \otimes  \set{\Delta}_m
  \oplus \set{\Delta}_n  \otimes {\set{L}_{\bvec{e}}}_{\, m}\\
 & \oplus
 \set{\Delta}_n  \otimes \set{\Delta}_m = \\
 =&\set{\Delta}_{m + n} \cap \set{\Delta}_{\sigma(m + n)}  
    \end{aligned}
\end{align}
where $\sigma(m + n)$
is the permutation that exchanges
the the $m$ comb with the $n$ comb, for example
\begin{align*}
 &  \set{\Delta}_{2 \otimes 1} = \set{\Delta}_{2 + 1} \cap
  \set{\Delta}_{\sigma(2 + 1)} = \\[8pt]
&=  \begin{aligned}
  \Qcircuit @C=0.6em @R=0.2em {
  \ustick{\scriptstyle{1}}&
                              \multigate{1}{\;}&
                                                 \ustick{\scriptstyle{2}}\qw&
  &
    \ustick{\scriptstyle{3}}&
                                \multigate{1}{\;}&
                                                   \ustick{\scriptstyle{4}}\qw&
  &
    \ustick{\scriptstyle{5}}&
                                \multigate{1}{\;}&
                                                   \ustick{\scriptstyle{6}}\qw\\
                            &
                              \pureghost{\;}&
                                              \qw&
                                                   \qw&
                                                        \qw&
                                                             \ghost{\;}&
                                                                         \qw&
                                                                              \qw&
                                                                                   \qw          &
                                                                                                  \ghost{\;}&}
                                                                                                \\[-11pt]
                                                                                                \underbrace{\;\quad\qquad\qquad}
                                                                                                \qquad \!
                                                                                                \underbrace{}
                                                                                                \:\,
                                                                                                \\[-4pt]
                                                                                                \bvec{2}
                                                                                                \qquad
                                                                                                \qquad \quad
                                                                                                \bvec{1} \quad
    \end{aligned}
                                                                                                             \;
                                                                                                             {\cap}
  \;
                                                                                                              \begin{aligned}
                                                                                                              \Qcircuit @C=0.6em @R=0.2em {
                                                                                                              \ustick{\scriptstyle{5}}&
                                                                                                                                          \multigate{1}{\;}&
                                                                                                                                                             \ustick{\scriptstyle{6}}\qw&
                                                      &
                                                        \ustick{\scriptstyle{1}}&
                                                                                    \multigate{1}{\;}&
                                                                                                       \ustick{\scriptstyle{2}}\qw&
                                                                                 &
                                                                                   \ustick{\scriptstyle{3}}&
                                                                                                               \multigate{1}{\;}&
                                                                                                                                  \ustick{\scriptstyle{4}}\qw\\
                            &
                              \pureghost{\;}&
                                              \qw&
                                                   \qw&
                                                        \qw&
                                                             \ghost{\;}&
                                                                         \qw&
                                                                              \qw&
                                                                                   \qw          &
                                                                                   \ghost{\;}&}
                                                                                 \\[-11pt]
                                                                                 \underbrace{}
                                                                                 \qquad \!
                                                                                 \underbrace{\;\quad\qquad\qquad}
                                                                                                \:\,
                                                                                                \\[-4pt]
                                                                                                \bvec{1}
                                                                                                \qquad
                                                                                                \qquad \quad
                                                                                                \bvec{2}
                                                                                                \quad
                                                                                                \quad
                                                                                                \;
                                                                                                \; \,
                                                                                                              \end{aligned}\;.
\end{align*}
Moreover, it is easy to verify that
\begin{align}
\label{eq:lambdacombnor}
  \lambda_{m+n} = \lambda_{\sigma(m+n)},
\end{align}
which, together with Equation~\eqref{eq:combtensor2} gives
\begin{align}
  \Evd {\bvec{m} \otimes \bvec{n}} =
  \Evd {\bvec{m} + \bvec{n}} \cap \Evd {\sigma (\bvec{m} + \bvec{n})}. 
\end{align}
Finally, let us consider the type $\bvec{n} \to \bvec{m}$.
By definition we have
$\bvec{n} \to \bvec{m} = \bvec{n} \to (\bvec{m} \to \bvec{1})$
and from Proposition~\ref{prp:curryingident} we have
$\bvec{n} \to \bvec{m} \equiv (\bvec{n} \otimes \bvec{m-1}) \to
\bvec{1}$.
Then, from Equation~\eqref{eq:eledetchar2} together with Equation~\eqref{eq:combtensor2}, we have
\begin{align}
  \label{eq:combtocomb}
  &
    \begin{aligned}
    \set{\Delta}_{n \to m} = & \set{Herm}(\hilb{H}_{n \otimes (m-1)}) \otimes
  \set{\Delta}_1 \\
  &\oplus \overline{\set{\Delta}}_{n \otimes
  (m-1)}\otimes(\set{L}_{\bvec{e}\,1} \oplus
\overline{\set{\Delta}}_{1}) =\\
=&\spn( \set{\Delta}_{(n + (m-1)) \to 1 } \cup
\set{\Delta}_{\sigma (n + (m-1)) \to 1 } )  
\end{aligned}
  \\
  &
    \begin{aligned}
      \implies
                               &\Evd {\bvec{n} \to \bvec{m}} = \\
=           &                    \set{Aff} \{
                               \Evd {(\bvec{n} + (\bvec{m-1})) \to\bvec{1} }
                               \cup\\
                               &
                               \Evd {\sigma (\bvec{n}
                               +(\bvec{m-1})) \to \bvec{1}}
                               \}
    \end{aligned} \nonumber
\end{align}
where in the last step we used Equation~\eqref{eq:lambdacombnor}.

\section{The inverse characterization problem}\label{sec:inverse}

In the previous section we studied the following problem:
given a type $x$ characterize the convex set $\Evd x$
of deterministic events of type $x$.
From Proposition~\ref{prop:char-high-order} we have
that the solution to this problem amounts to the evalution
of the function
\begin{align}
  \label{eq:charamap}
  \begin{aligned}
      \Upsilon: \set{Types} &\to \mathbb{R} \times
 \set{S}(\set{Traceless}(\hilb{H}_x))  \\
  x &\mapsto
  \begin{pmatrix}
    \Upsilon_1(x) = \lambda_x\\
    \Upsilon_2(x) = \set{\Delta}_x
  \end{pmatrix}
  \end{aligned}
\end{align}
where $\set{S}(\set{Traceless} (\hilb{H}_x))$ denotes the set of real
subspaces of $\set{Traceless} (\hilb{H}_x)$.  Both $\Upsilon_1$ and
$\Upsilon_2$ can be evaluated by recursively applying
Equations~\eqref{eq:lambdarecursive} and~\eqref{eq:eledetchar2}.  All
the relevant information about higher-order quantum theory is
encoded in the map $\Upsilon$ and in the cone of positive operators.
For example, let us consider the set $\Upsilon_2(\set{Types})$,
i.e. the range of the map $\Upsilon_2$. This set contains
a relevant information about the mathematical structure of quantum theory,
namely what are the
linear subspaces that are relevant in higher-order quantum theory.
From this point of view, it is obvious that the set of quantum transformations and
higher-order maps exhibits a much richer structure than
the set of normalized quantum states, which are simply all the  positive
operators with unit trace.

For example, one
could wonder what is
the image under the map $\Upsilon_2$
of the set of types which have the same Hilbert space
(i.e. $\hilb{H}_x=\hilb{H}_y = \hilb{H}$
for two types $x$ and $y$).
More generally, we can
address the following question:\\ $\,$\\
{\bf Inverse characterization problem:} \emph{Given an Hilbert space $\hilb{H}$
and a linear subspace $\set{\Delta} \subseteq
\set{Traceless}(\hilb{H})$, 
which are the a types $x$ such that
$\hilb{H}_x = \hilb{H}$ and $\set{\Delta} = \set{\Delta}_x$ (if any)?}
\\
$\,$\\
Roughly speaking, the inverse characterization problem amounts to computing
the inverse map ${\Upsilon_2}^{-1}$. This is a much harder task
than the direct one.
We now address an instance of this problem,
which we find particularly instructive.

Let $\hilb{H}:= \mathbb{C}^2 \otimes  \mathbb{C}^2$
and  $\set{\Delta} := \set{Traceless}(\mathbb{C}^2) \otimes
\set{Traceless}(\mathbb{C}^2)$ and let us suppose that there exists a type
$z$ such that
$\hilb{H}_z = \hilb{H}$ and $\set{\Delta} = \set{\Delta}_z$.
First we notice that $z$ cannot be an elementary type.
If $z=A$ with $\hilb{H}_A = \hilb{H}$ it must be
$ \set{\Delta}_A = \set{Traceless}(\hilb{H}) $
and $\dim (\set{Traceless}(\hilb{H})) = 15$ while
$\dim (\set{\Delta} ) =9$.
Let us then  suppose that
$z = x \to y $.
Since $\dim(\hilb{H}_z) = 4$
we must   have
$\dim(\hilb{H}_x) \dim(\hilb{H}_y) =4  $.
Moreover, 
since $I \to y \equiv y$
we suppose that $ \dim(\hilb{H}_x) > 1 $.
We have therefore the following two possibilities:
\begin{align*}
 &\dim(\hilb{H}_x)= 4 \mbox{ and }  \dim(\hilb{H}_y) = 1 \mbox{ or }\\
  &\dim(\hilb{H}_x)= 2 \mbox{ and } \dim(\hilb{H}_y) = 2.
\end{align*}
From Equation \eqref{eq:eledetchar2} we have
\begin{align}
 &\set{\Delta}_z  = [\set{Herm}(\hilb{H}_y) \otimes  \overline{\set{\Delta}_{x}}]
    \oplus
[{\set{\Delta}_{y}}   \otimes ( \set{L}_{\bvec{e}}  \oplus
  \set{\Delta}_{x} )] \implies \nonumber\\
&9 =  \dim(\set{\Delta}_z )
   =d_y^2(d_x^2 -1-a_x ) +
  a_y(1+a_x)  \label{eq:dimensionconstraint}\\
  \nonumber
\end{align}
where $d_x = \dim(\hilb{H}_x)$,
$d_y = \dim(\hilb{H}_y)$,
$a_x = \dim(\set{\Delta}_{x} ) < d^2_x$
and
$a_y = \dim(\set{\Delta}_{y} )<d^2_y$.
If we assume  $d_y=1$ and $d_x=4$, i.e. $z \equiv x \to I$,
then we must have $a_x=6$.
Since for any elementary type $E$ we must have
$a_E = d_E^2-1$, the type $x$ cannot be elementary.
Then there must exist $f \neq I $ and $g$ such that $x = f \to g $.
Since $\overline{\overline{x}} = x$ we must have that
$g \neq I $,  in order to avoid the tautology $z \equiv (z \to I) \to I$.
Then, since $d_x=4$, we must have $d_f = d_g =2$ and then
\begin{align*}
  6 =\dim(\set\Delta_{f\to g})=4(4-1-a_f) + a_g(1+a_f)
\end{align*}
which cannot be satisfied for any couple $a_x,a_y$
such that $0 \leq a_f,a_g \leq 3$.
Therefore the case $d_x= 4$, $d_y = 1 $
must be discarded.
Let us then consider the case
$d_x = d_y =2$.
Eq.~\eqref{eq:dimensionconstraint} gives
\begin{align*}
  9 = 4(4-1-a_x) + a_y(1+a_x)
\end{align*}
which  cannot be satisfied for any couple $a_x,a_y$
such that $ 0 \leq a_x,a_y \leq 3 $.

This result shows that, given $\hilb{H}$ and
$\set{\Delta} \subseteq \set{Traceless}(\hilb{H})$, it might be the
case that there exists no type such that $\hilb{H}_x = \hilb{H}$ and
$\set{\Delta} = \set{\Delta}_x$. Notice that this no-go result holds
also in the simplified scenario where $\hilb H$ is specified from the
beginning as the tensor product of elementary type spaces.  Therefore,
for a given Hilbert space, the characterization of the set of
subspaces $\set{\Delta}\subseteq \set{Traceless}(\hilb H)$ which
correspond to some type is far from trivial.  In comparison to the
first item in the notion of admissibility, that reduces to the
positivity requirement, the second one, which involves the notion of
deterministic, entails a much more complex mathematical structure.

\section{Conclusions}\label{sec:conclu}
We formulated a fully operational framework for higher-order quantum
theory based on a set of axioms regarding the notion of
admissible transformation. This definition is recursive, and 
requires a type system in the first place, allowing for the 
labelling of sets of transformations,
basically through their common domain and range. This structure is
shared with classical {\em typed lambda calculus}
\cite{selinger2008lecture}, where the typing rules are necessary to
select well-formed expressions. We provide a recursive
characterization of maps of an arbitrary type, which is then used to
prove a set of basic type equivalences.
                                                      
    

 {Although some similarities, it is worth stressing that
  our framework fundamentally differs from the works on
  denotational semantics for a quantum programming language
  and quantum lambda
  calculus\cite{selinger2004towards,selinger2009quantum,10.1007/11417170_26,hasuo2017semantics,malherbe2013categorical,Pagani:2014:AQS:2578855.2535879}.
  In particular, one of the goals of our approach is to encompass
  quantum computation without a definite causal order. For
  example, the quantum SWITCH map, which we previously
  described, is a paradigmatic example of a higher order map
  which our formalism can describe, but that lies outside
  the framework of Ref\cite{10.1007/11417170_26}, as first
  noticed in Ref.\cite{PhysRevA.88.022318}.  A categorical
  framework closely related to the one presented in this
  contriburion, has been presented by Kissinger and Uijlen
  in Ref.\cite{8005095}.  They introduce a categorical
  construction which sends certain compact closed categories
  $\mathcal{C}$ to a new category
  $\mathrm{Caus}[\mathcal{C}]$.  This procedure can be
  applied to Selinger's CPM construction of
  Ref.\cite{SELINGER2007139}, which does not take
  normalization, and hence causality, into account.  On one
  hand, by combining this two results, one obtains the
  hierarchy of higher order quantum maps of our framework.
  On the other hand, from a foundational perspective,
  assuming CPM's construction amounts to assume complete
  positivity for all the maps in the hierarchy without any
  physical motivation.  Moreover, several assumptions of the
  framework in Ref.\cite{8005095}, for example that
  second-order causal processes factorise, are also not
  operationally justified.  The main goal of our work is to
  give a fully operational (i.e. avoiding explicit reference
  to the mathematical properties of maps in the hierarchy)
  formulation of higher order quantum theory
  which can
  encompass indefinite causal structures.  In particular, we
  gave an operational definition of admissibility which does
  not assume complete positivity.  In our setting, the proof
  that any positive operator (up to suitable and necessary
  rescaling) is an admissible higher order map is
  nontrivial.  On the other hand, by assuming CPM and
  Ref.\cite{8005095} construction, this same result becomes
  a rather straightforward observation.}

 {Higher-order quantum theory must be
thought of as an extension of quantum theory,
which provides
a natural unfolding of a part of the theory that is
implicitly contained in any of its formulations. As such,
higher-order quantum theory has a fundamental value, being a
new standpoint for the analysis of the peculiarities of
quantum theory.  The axioms of our framework have a purely operational nature and do not rely
on the specific mathematical structure of quantum
theory. Therefore, with proper care, our framework can be
applied to general probabilistic theories.  In particular
the most important ingredient we used is the Choi
isomorphism, that can be always provided in theories where
local discriminability holds. If the latter does not hold
one must reformulate the recursive definition of admissible
events avoiding the Choi correspondence. In this case, since
parallel composition is not simply translated in the tensor
product rule, a transformation is not simply a single
matrix, but a possibly infinite family of matrices
representing the action of the map on all possible extended
systems.}

 {The framework that we introduced leads to several open problems.
An interesting question is to
determine what types, if any, can
be attributed to a given subspace of linear maps. An even harder
problem is to determine all the possible types of a given linear map.}

 {In this work, we proved a family of equivalences between types of higher order maps. Therefore, another question that naturally arises is whether there exists a
complete set of type equivalences,
i.e. a set of type equivalences
such that their compositions provide an alternative
characaterization
of the hierarchy of higher order quantum maps.}
 {Moreover, following the case of causally ordered quantum
networks, one would like to infer the causal structure
of an higher order map from its type.}

 {Finally, the present work only partially addresses
the composition of types.
It is implicit in our definition that,
given a map of type $x$ and a
map of type $x \to y$,
they can be composed and give a map of type $y$.
However, our formalism does not provide any
formal rules which
would translate a partial application of a higher order map
(apart from the easiest
case of the extension with an elementary type).
In order to have a theory of computation,
a comprehensive set
of rules that encompasses all the admissible
composition of maps must be given.}

\appendix

\section{Proof of Theorem \ref{prop:positiveconeadmis}}
\label{sec:proof-theor-refpr}
\begin{lemma} 
$X\in\Eva x$ if and only if it satisfies item \ref{item1} of definition~\ref{def:admissiblevents}, and there exist $\{X_i\}_{i=1}^n\subseteq\Eva x$ such that $X+\sum_{i=1}^nX_i\in\Evd x$.
\label{lem:sticazzetti}
\end{lemma}

\Proof The proof proceeds by induction. The statement is straightforwardly true for $x\in\ETy$. Let it
now be true for any $y$ and $y \para E'$, for arbitrary $E'\in\ETy$ and $y \preceq x$.
We need to prove that the statement holds for $x \para E$ for any
arbitrary $E\in\ETy$. If $x$ is not elementary we can write
$x \para E = (y \to z )\para E =y \to z \para E$ for some $y,z \preceq x$.

Clearly, if $X\in\Eva {x \para E} $, by definition \ref{def:admissiblevents}
there must exist $\{X_i\}_{i=1}^n\subseteq\Eva{x\para E}$ such that, upon defining $X_0:=X$ and 
$D:=\sum_{i=0}^nX_i$, one has $[\Ch^{-1}(D)\otimes\tI_{E'}](\Evd{y\para E'})\subseteq\Evd {z \para EE'}$. 
Now, for $Y_0\in\Eva{y\para E'}$, there exist $\{N_j\}_{j=1}^k\subseteq\Eva{y\para E'}$ such that, 
by the induction hypothesis, $G:=Y_0+\sum_{j=1}^k N_j\in\Evd{y\para E'}$. Thus,
\begin{align*}
[\Ch^{-1}(D)&\otimes\tI_{E'}](G)\\
&=\sum_{j=0}^k[\Ch^{-1}(D)\otimes\tI_{E'}](Y_j)\in\Evd{z\para EE'},
\end{align*}
which means that $[\Ch^{-1}(D)\otimes \tI_E](Y_0)$ is admissible, again using the induction hypothesis. 
Then $D$ satisfies item \ref{item1} of definition~\ref{def:admissiblevents}.

The proof of the converse statement is trivial.
\qed

\begin{lemma} 
If $X,X'\in\Eva x$, then $X+X'\in\Eva x$ if and only if there exist $\{X_i\}_{i=1}^n\subseteq\Eva x$ such that $X+X'+\sum_{i=1}^nX_i\in\Evd x$.
\label{lem:sticazzi}
\end{lemma}

\Proof The direct statement can be proved by the same technique as for lemma~\ref{lem:sticazzetti}.
Now, for the converse, we proceed by induction. Suppose that the statement holds for $y$ and $y\para E$, 
for every $y\prec x$ and $E\in\ETy$. Suppose now that $X,X'\in\Eva{x\para E}$, and that $\{X_i\}_{i=1}^n$ exists such that
$D:=X+X'+\sum_{i=1}^nX_i\in\Evd{ x \para E}$. Since $\Ch^{-1}(X)$ and
$\Ch^{-1}(X')$ satisfy item \ref{item1}, then for every
$Y\in\Eva{y\parallel E'}$, both $[\Ch^{-1}(X)\otimes\mathcal{I}_{E'}](Y)$ and
$[\Ch^{-1}(X)\otimes\mathcal{I}_{E'}](Y)$ are in $\Eva{(z\parallel E)\para
  E'}= \Eva{z\para EE'}$. Moreover,
there are $\{Y_i\}_{i=1}^n$ such that
$Y_0:=Y+\sum_{j=1}^mY_i\in\Evd{y\parallel E'}$, and thus
$[\Ch^{-1}(D)\otimes \trasf{I} _E](Y_0)\in\Evd{z\parallel EE'}$. On the other
hand,
\begin{align*}
  [&\Ch^{-1}(D)\otimes\tI_{E'}](Y_0)\\
   &=[\Ch^{-1}(X)\otimes\tI_{E'}](Y)+[\Ch^{-1}(X')\otimes\tI_{E'}](Y)\\
   &+\sum_{i=1}^n[\Ch^{-1}(X_i)\otimes\tI_{E'}](Y)+\sum_{j=1}^m[\Ch^{-1}(D)\otimes\tI_{E'}](Y_j).
\end{align*}
By the induction hypothesis, the above condition assures us that
$[\Ch^{-1}(X)\otimes\tI_{E'}](Y)+
[\Ch^{-1}(X')\otimes\tI_{E'}](Y)\in\Eva{z\parallel EE'}$, and thus
$[\Ch^{-1}(X+X')\otimes\tI_{E'}](Y)\in\Eva{z\parallel EE'}$. This implies
that $\Ch^{-1}(X+X')\otimes\tI_{E'}$ maps $\Eva{y\parallel E'}$ into
$\Eva{z\parallel EE'}$. Thus, all the requirements of definition
\ref{def:admissiblevents} are fulfilled by $X+X'$, and
$X+X'\in\Eva{ x \para E}$ for any $E$.\qed

\begin{lemma} $X\in\Ev x$ is admissible if and only if it satisfies
  item \ref{item1} of definition \ref{def:admissiblevents} and there
  exists $X'$ satisfying item \ref{item1} such that $X+X'\in\Evd x$.
\label{lem:oralofaccio}
\end{lemma}
\Proof Let $X$ and $X'$ satisfy item \ref{item1} of definition
\ref{def:admissiblevents} and $X+X'\in\Evd x$. Then $X,X'\in\Eva
x$. Viceversa, if $X$ is admissible then it satisfies item \ref{item1}
and there exist $\{X_i\}_{i=1}^n$ such that, for every $1\leq i\leq n$,
$X_i$ satisfies item \ref{item1}, and $X+\sum_{i=1}^n\in\Evd x$. Thus,
for every $1\leq i\leq n$ it is $X_i\in\Eva x$. By iterating lemma
\ref{lem:sticazzi}, we have $S:=\sum_{i=1}^nX_i\in\Eva x$. Moreover, clearly
$X+S\in\Evd x$.\qed

\begin{corollary}
$X\in\Ev x$ is admissible if and only if it satisfies item \ref{item1} of definition \ref{def:admissiblevents} and there exists $X'\in\Eva x$ such that $X+X'\in\Evd x$.
\label{cor:embeh}
\end{corollary}

\begin{lemma}
  Let $X\in\Eva{x}$, and $\rho\in\Eva E$. Then
  $X\otimes\rho\in\Eva{x\parallel E}$. Moreover, if $X\in\Evd{x}$ and
  $\rho\in\Evd{E}$, then $X\otimes\rho\in\Evd{x\parallel E}$.
\label{lem:vedisopra}
\end{lemma}
\Proof Also in this case we proceed by induction. The statement is
true for $x\in\ETy$. Let now the statement be true for $y \para F'$
for any $y \preceq x$ and arbitrary $F'\in\ETy$.  Since $x$ is not an
elementary type, we have $x \para F = y \to z \para F $ for some
$y,z \preceq x$.  Let $X\in\Eva{x \para F}$. Let
$R\in\Eva{y\parallel F'}$. Then
$[\Ch^{-1}(X\otimes\rho)\otimes\tI_{F'}](R)=[\Ch^{-1}(X)\otimes\tI_{F'}](R)\otimes\rho$,
and since by definition
$[\Ch^{-1}(X)\otimes\tI_{F'}](R)\in\Eva{z\parallel FF'}$, we have by
the induction hypothesis that
$[\Ch^{-1}(X\otimes\rho)\otimes\tI_{F'}](R)=[\Ch^{-1}(X)\otimes\tI_{F'}](R)\otimes\rho\in\Eva{z\parallel
  FF'E}$.  Thus,
$[\Ch^{-1}(X\otimes\rho)\otimes\tI_{F'}][\Eva{y\parallel
  F'}]\subseteq\Eva{z\parallel FF'E}$ for every $F'\in\ETy$. Now, if
$X$ is admissible, then by lemma \ref{lem:oralofaccio}, there is $X'$
such that $\Ch^{-1}(D)$ is a deterministic map of type
$x=y\to z\para F$, where $D:=X+X'$, and
$[\Ch^{-1}(X')\otimes\tI_{F'}][\Eva{y\parallel
  F'}]\subseteq\Eva{z\parallel FF'}$ for every $F'\in\ETy$. Similarly,
there is $\sigma\in\Evd{E}$ such that $\sigma\geq\rho$. Thus,
$[\Ch^{-1}(D\otimes\sigma)\otimes\tI_{F'}](Y)=[\Ch^{-1}(D)\otimes\tI_{F'}](Y)\otimes\sigma$
which is deterministic by the induction hypothesis, and
$D\otimes\sigma=X\otimes\rho+X'\otimes\rho+X\otimes\tau+X'\otimes\tau$,
where $\tau:=\sigma-\rho\geq 0$. Thus $X\otimes\rho$ is
admissible. As to the second item in the thesis, if $X\in\Evd {x \para F} $, $\rho\in\Evd E $, and
$Y\in\Evd{y\parallel F'}$, then
$[\Ch^{-1}(X)\otimes\tI_{F'}](Y)\in\Evd{z\parallel FF'}$ and thus, by the
induction hypothesis,
$[\Ch^{-1}(X\otimes\rho)\otimes\tI_{F'}](Y)\in\Evd{z\parallel FF'E}$, and
thus
$[\Ch^{-1}(X\otimes\rho)\otimes\tI_{F'}][\Evd{y\parallel
  F'}]\subseteq\Evd{z\parallel FF'E}$. \qed
$\,$\\

We now prove the following crucial lemma. Let $\Evp x$ denote the set $\{P\in\Ev x\mid\exists \lambda\geq0,\ P'\in\Eva x:\ P=\lambda P'\}$.

\begin{lemma} 
For every type $x\in\Ty$, the set $\Evp x$ is the full positive cone in $\Lin{\hilb H_x}$.
\label{lem:positcone}
\end{lemma}
\Proof
Let us restate the hypothesis as follows. For every type
$x\in\Ty$, the sets $\Evp x$, $\Evp{x \to I}$, and
$\Evp {x\parallel E}$ are the full positive cones in
$\Lin{\hilb H_x}$, $\Lin{\hilb H_x}$, and
$\Lin{\hilb H_x\otimes\hilb{H}_E}$, respectively. This new form of the
thesis is amenable to a proof by induction as follows.
The thesis holds for elementary types.
Let now
$x=y_1\to y_2$, and let us suppose that the thesis holds for
$y_1$ and $y_2$. In the first place, this implies that a necessary
condition for $M$ to be admissible is that $M$ is the Choi of a
completely positive map, and thus it must be positive. We have then
that the set $T_+(x) $ is contained in the cone of positive
operators in $\mathcal{L}(H_x)$. Moreover, the
induction hypothesis implies that there exist full-rank elements of
type $\Evd{{y_1 \to I}}$ and $\Evd{y_2}$. Let $\overline{Y}_1$ and
$Y_2$ denote two such elements. We now claim that
$X:=\overline{Y}_1 \otimes Y_2$ is proportional to an admissible element
of type $x$. Indeed, let $Y_{1E}$ denote an arbitrary admissible
element of type $y_1\parallel E$. One can easily check that
$[\Ch^{-1}(X)](Y_{1E})=\rho_E\otimes Y_2$, where
$\rho_E:=[\Ch^{-1}(\overline Y_1)](Y_{1E})$. Now, by lemma
\ref{lem:vedisopra}, $\rho_E\otimes Y_2$ is admissible. Thus, $\Ch^{-1}(X)$ maps
admissible elements of $\Eva{y_1\parallel E}$ to admissible elements
of $\Eva{y_2\parallel E}$. Moreover,
$\Ch^{-1}(\overline Y_1)\otimes\tI_E$ maps deterministic elements of
$\Evd{y_1\parallel E}$ to elements of the form $\rho_E\otimes Y_2$,
with $\rho_E\in\Evd{E}$ and $Y_2\in\Evd{y_2}$. By lemma \ref{lem:vedisopra} these elements
are in $\Evd{y_2\parallel E}$. Finally, we proved that
$X=Y_2\otimes \overline Y_1$ satisfies the conditions for an
admissible element of $\Evd{y_1\to y_2}$, and it is full-rank. Exactly
the same argument can be used to prove the statement for
$x\parallel F=y_1\to y_2\parallel F$. Finally, for the case
$x={(y_1\to y_2)\to I}$ one can easily check that
$\overline Y_2\otimes Y_1$ is in $\Ev{{(y_1\to
    y_2) \to I }}$. Moreover, it is admissible since
$[\Ch^{-1}(\overline Y_2\otimes Y_1)\otimes \tI_E](X)=X_E$ corresponds
to the application of $\Ch^{-1}(\overline Y_2)$ to
$[\Ch^{-1}(X)](Y_1)$. Now, since $\Ch^{-1}(X)$ is admissible by
hypothesis as well as $\Ch^{-1}(Y_1)$ and $\Ch^{-1}(\overline Y_2)$,
what we get in the end is an admissible element of $\Eva E$. Moreover,
$\overline Y_2\otimes Y_1$ is deterministic, since
$[\Ch^{-1}(X)](Y_1)$ is deterministic for deterministic $X$, and thus
we have that
$X_E=[\Ch^{-1}(\overline Y_2\otimes Y_1)\otimes \tI_E](X)$ is
deterministic for any deterministic $X$. Moreover,
$\overline Y_2\otimes Y_1$ is full rank.
Now, let $M \geq 0$ be a
positive operator on $\hilb{H}_x$.
Then, since $ \overline{Y}_1 \otimes Y_2$ is positive and full-rank,
there exist a positive coefficient $\lambda$ such that
$\lambda M \leq \overline{Y}_1 \otimes Y_2$.
One can then easily verify that
$\lambda M $ and  $ \overline{Y}_1 \otimes Y_2 - \lambda M $
satisfy the hypotheses of Definition~\ref{def:admissiblevents}.
Therefore the positive cone in $\mathcal{L}(\hilb{H}_x)$
is contained in $T_+(x)$. The same result can be proved for ${x \to I}$ and $x \para E$.
\qed

$\,$

We are now ready to prove  theorem \ref{prop:positiveconeadmis}

\Proof
By lemma \ref{lem:positcone}, if $M\in\Eva x$ then $M\geq
0$. Moreover, by corollary \ref{cor:embeh} there exists $M'\geq0$ such
that $D:=M+M'\in\Evd x$. Thus, $M'=D-M\geq0 \implies M \leq D $.

To prove the converse, let us consider a non elementary type $x \para E$, i.e.
$x=y\to z \para E$, for an arbitrary elementary type $E$ (the thesis is trivially true
if $x$ is elementary).
Let us suppose that the thesis holds for any $y \para E$, $y\preceq x$
and consider $0\leq M \leq D\in\Evd {x \para E }$, $N:= D-M$.
From the induction hypothesis,  one can verify that $M$ and $N$ satisfy the hypothesis of
Definition~\ref{def:admissiblevents} and therefore $M$ is an admisible
event of type $x \para E$ for arbitrary $E$. \qed

\section{Proof of Theorem~\ref{prop:deteventmadeeasy}}
\label{sec:proof-prop-deteasy}
The proof relies on the following preliminary results.
\begin{lemma}
  \label{lmm:superpositive}
  For arbitrary $x$ and $y$,
consider the type $x\to y$.
Let  $R\in \Ev {x \to y} $ such that $R \geq 0$
  and, for all $E$,
  $[\Ch^{-1}(R)\otimes \mathcal{I}_E](\Evd {x\para E}) \subseteq
  \Evd {y\para E}$. Then
  $R \in \Evd {x\to y}$ 
\end{lemma}
\Proof
We first need to show that
 $[\Ch^{-1}(R)\otimes \mathcal{I}_E](\Eva {x\para E}) \subseteq
  \Eva {y\para E}$.
  Let us fix an arbitrary $O_{x \para E}\in \Eva {x\para E}$.
  From Theorem~\ref{prop:positiveconeadmis}
  we have that
  $0 \leq O_{x \para E} \leq D_{x \para E} $
  for some $D_{x \para E}  \in \Evd {x\para E}$.
Since 
$R \geq 0$, the map
$\Ch^{-1}(R)$ is completely positive.
Therefore we have $  [\Ch^{-1}(R)\otimes  \mathcal{I}_E](D_{x \para E}-O_{x \para E}) \geq
  0 $, which implies
\begin{align*}
  0& \leq [\Ch^{-1}(R)\otimes \mathcal{I}_E](O_{x \para E}) \\
              &\leq (
                [\Ch^{-1}(R)\otimes  \mathcal{I}_E] ( D_{x \para E}) \in
  \Evd {y\para E}.
\end{align*}
We conclude that $[\Ch^{-1}(R)\otimes{\mathcal I}_E](O_{x \para E}) \in \Eva {y\para E}$.
Finally, using the hypothesis that $[\Ch^{-1}(R)\otimes{\mathcal I}_E](\Evd {x\para E}) \subseteq \Evd {y\para E}$, by theorem \ref{prop:positiveconeadmis} the thesis follows. \qed
\begin{lemma}\label{lem:necesuffpart}
For every $E,E'\in\ETy$, and for every $R \in \Eva {x \para EE'}$ one has that 
 \begin{align}
    \label{eq:partialtracedet}
  R \in \Evd {x \para EE'} \iff \Tr_E[R] \in \Evd {x\para E'}  
    \end{align}
\end{lemma}
\Proof
The proof is by induction.
For $x\in\ETy$ the thesis is easily verified.
Let us suppose that the thesis holds for any $y \preceq x$, 
and let us write $x = y\to z$. Clearly, since
$R \in \Eva {x \para EE'} $, by lemma \ref{lem:positcone} 
we have that $R \geq 0$. By lemma \ref{lmm:superpositive} a necessary and sufficient condition for $R\in\Evd{y\to z\para EE'}$ is then that
$[\Ch^{-1}(R)\otimes\tI_F](\Evd{y\para F})\subseteq \Evd{z\para EE'F}$.
Let us now fix an arbitrary $F$ and an arbitrary $D \in \Evd {y \para F}$. 
By the induction hypothesis, a necessary and sufficient condition for $[\Ch^{-1}(R)\otimes\tI_{F}](D)$ to be
in $\Evd{z\para EE'F}$ is that
\begin{align*}
\Tr_{F}[(\Ch^{-1}(R)\otimes\tI_{F})(D)]\in\Evd{z\para EE'}.
\end{align*}
However, we have
\begin{align*}
\Tr_{F}[\Ch^{-1}(R)\otimes\tI_{F}](D)]=[\Ch^{-1}(R)](\Tr_{F}[D]).
\end{align*}
We can now rewrite the necessary and sufficient condition for 
$R\in\Evd{y\to z\para EE'}$ as
\begin{align*}
\Tr_{yF}[({D}^T\otimes I_{z\para EE'})(R\otimes I_F)]\in\Evd{z\para EE'}.
\end{align*}
By the induction hypothesis, again this is equivalent to
\begin{align*}
\Tr_{yEF}=[({D}^T\otimes I_{z\para EE'})(R\otimes I_F)]\in\Evd{z\para E'},
\end{align*}
namely 
\begin{align*}
&\Tr_{yF}[({D}^T\otimes I_{z\para E'})(\Tr_E[R]\otimes I_F)]\\
&=\Ch^{-1}(\Tr_E[R])(\Tr_F[D])\\
&=\Tr_F[(\Ch^{-1}(\Tr_E[R])\otimes\tI_F)(D)]\in\Evd{z\para E'},
\end{align*}
which, again by the induction hypothesis, is equivalent to $(\Ch^{-1}(\Tr_E[R])\otimes\tI_F)(D)\in\Evd{z\para E'F}$.
If this holds for arbitrary $F$ and arbitrary $D\in\Evd{y\para F}$, the above condition is finally equivalent to
condition \eqref{eq:partialtracedet}. \qed

Let us now address the proof of Theorem~\ref{prop:deteventmadeeasy}.
\Proof By lemma \ref{lem:positcone} and lemma \ref{lmm:superpositive}, a necessary and sufficient condition for $M\in\Evd{x\to y}$ is that $M\geq0$ and 
$[\Ch^{-1}(M)\otimes\tI_E](\Evd{x\para E})\subseteq\Evd{y\para E}$. Now, for $M\geq 0$, by lemma \ref{lem:necesuffpart} the above necessary and sufficient condition is equivalent to the requirement that for every $D\in\Evd{x\para E}$ one has
\begin{align*}
\Tr_E[(\Ch^{-1}(M)\otimes\tI_E)(D)]=\Ch^{-1}(M)(\Tr_E[D])\in\Evd{y}.
\end{align*}
However, for this condition to hold it is necessary and sufficient that $\Ch^{-1}(M)(\Evd{x})\subseteq\Evd{y}$.\,\qed

  \vskip6pt

\enlargethispage{20pt}


\dataccess{This article has no additional data.}

\aucontribute{All the authors equally contributed to the present manuscript.}

\competing{The authors declare that there are no competing interests.}

\funding{This publication was made possible through the support of a
  grant from the John Templeton Foundation under the project ID\#
  60609 Causal Quantum Structures. The opinions expressed in this
  publication are those of the authors and do not necessarily reflect
  the views of the John Templeton Foundation.}

\ack{The authors are grateful to the anonymous referees who reviewed
  and commented on the earlier submission of this paper.}



\bibliographystyle{RS}
\bibliography{bibliography}

\begin{thebibliography}{99}

\bibitem{Kraus}
Kraus K. 1983 {\em States, effects and operations: fundamental notions of
  quantum theory}.
Springer.

\bibitem{chiribella2008transforming}
Chiribella G, D'Ariano GM, Perinotti P. 2008  Transforming quantum operations:
  Quantum supermaps. {\em EPL (Europhysics Letters)} \textbf{83}, 30004.

\bibitem{1402-4896-2014-T163-014013}
D'Ariano GM, Manessi F, Perinotti P. 2014  Determinism without causality. {\em
  Physica Scripta} \textbf{2014}, 014013.

\bibitem{baumeler2014perfect}
Baumeler A, Wolf S. 2014  Perfect signaling among three parties violating
  predefined causal order. In {\em Information Theory (ISIT), 2014 IEEE
  International Symposium on} pp. 526--530. IEEE.

\bibitem{selinger_valiron_2009}
Selinger P, Valiron B. 2009 pp. 135--172.
In {\em Quantum Lambda Calculus}, pp. 135--172. Cambridge University Press.

\bibitem{PhysRevLett.101.060401}
Chiribella G, D'Ariano GM, Perinotti P. 2008  Quantum Circuit Architecture.
  {\em Phys. Rev. Lett.} \textbf{101}, 060401.

\bibitem{PhysRevA.80.022339}
Chiribella G, D'Ariano GM, Perinotti P. 2009  Theoretical framework for quantum
  networks. {\em Phys. Rev. A} \textbf{80}, 022339.

\bibitem{bisio2016quantum}
Bisio A, Chiribella G, D'Ariano GM, Perinotti P. 2016  Quantum networks:
  general theory and applications. {\em Acta Physica Slovaca} \textbf{61}, 273.

\bibitem{Gutoski:2007:TGT:1250790.1250873}
Gutoski G, Watrous J. 2007  Toward a General Theory of Quantum Games. In {\em
  Proceedings of the Thirty-ninth Annual ACM Symposium on Theory of Computing}
  STOC '07 pp. 565--574 New York, NY, USA. ACM.

\bibitem{PhysRevA.88.022318}
Chiribella G, D'Ariano GM, Perinotti P, Valiron B. 2013  Quantum computations
  without definite causal structure. {\em Phys. Rev. A} \textbf{88}, 022318.

\bibitem{Oreshkov:2012aa}
Oreshkov O, Costa F, Brukner {\v{C}}. 2012  Quantum correlations with no causal
  order. {\em Nature communications} \textbf{3}, 1092.

\bibitem{Hardy:2007aa}
Hardy L. 2007  Towards quantum gravity: a framework for probabilistic theories
  with non-fixed causal structure. {\em Journal of Physics A: Mathematical and
  Theoretical} \textbf{40}, 3081.

\bibitem{PhysRevA.86.040301}
Chiribella G. 2012  Perfect discrimination of no-signalling channels via
  quantum superposition of causal structures. {\em Phys. Rev. A} \textbf{86},
  040301.

\bibitem{COLNAGHI20122940}
Colnaghi T, D'Ariano GM, Facchini S, Perinotti P. 2012  Quantum computation
  with programmable connections between gates. {\em Physics Letters A}
  \textbf{376}, 2940 -- 2943.

\bibitem{Facchini2015324}
Facchini S, Perdrix S. 2015  Quantum circuits for the unitary permutation
  problem. {\em Lecture Notes in Computer Science (including subseries Lecture
  Notes in Artificial Intelligence and Lecture Notes in Bioinformatics)}
  \textbf{9076}, 324--331.
cited By 0.

\bibitem{1367-2630-17-10-102001}
Ara{\'u}jo M, Branciard C, Costa F, Feix A, Giarmatzi C, Brukner {\v C}. 2015
  Witnessing causal nonseparability. {\em New Journal of Physics} \textbf{17},
  102001.

\bibitem{PhysRevX.8.011047}
Castro-Ruiz E, Giacomini F, Brukner icv. 2018  Dynamics of Quantum Causal
  Structures. {\em Phys. Rev. X} \textbf{8}, 011047.

\bibitem{Procopio:2015ab}
Procopio LM, Moqanaki A, Ara{\'u}jo M, Costa F, Alonso~Calafell I, Dowd EG,
  Hamel DR, Rozema LA, Brukner {\v C}, Walther P. 2015  Experimental
  superposition of orders of quantum gates. {\em Nature Communications}
  \textbf{6}, 7913 EP --.

\bibitem{PhysRevA.93.052321}
Rambo TM, Altepeter JB, Kumar P, D'Ariano GM. 2016  Functional quantum
  computing: An optical approach. {\em Phys. Rev. A} \textbf{93}, 052321.

\bibitem{Perinotti2017}
Perinotti P. 2017  Causal structures and the classification of higher order
  quantum computations. In {\em Time in Physics} pp. 103--127. Springer.

\bibitem{8005095}
Kissinger A, Uijlen S. 2017  A categorical semantics for causal structure. In
  {\em 2017 32nd Annual ACM/IEEE Symposium on Logic in Computer Science (LICS)}
  pp. 1--12.

\bibitem{Abramsky:1994ab}
Abramsky S, Jagadeesan R. 1994  Games and Full Completeness for Multiplicative
  Linear Logic. {\em The Journal of Symbolic Logic} \textbf{59}, 543--574.

\bibitem{coecke2017picturing}
Coecke B, Kissinger A. 2017 {\em Picturing Quantum Processes: A First Course in
  Quantum Theory and Diagrammatic Reasoning}.
Cambridge University Press.

\bibitem{PhysRevA.81.062348}
Chiribella G, D'Ariano GM, Perinotti P. 2010  Probabilistic theories with
  purification. {\em Phys. Rev. A} \textbf{81}, 062348.

\bibitem{chiribella2011informational}
Chiribella G, D'Ariano G, Perinotti P. 2011  Informational derivation of
  quantum theory. {\em Phys. Rev. A} \textbf{84}, 012311--012350.

\bibitem{d2017quantum}
D'Ariano GM, Chiribella G, Perinotti P. 2017 {\em Quantum theory from first
  principles: an informational approach}.
Cambridge University Press.

\bibitem{CHOI1975285}
Choi MD. 1975  Completely positive linear maps on complex matrices. {\em Linear
  Algebra and its Applications} \textbf{10}, 285 -- 290.

\bibitem{d2014feynman}
D'Ariano GM, Manessi F, Perinotti P, Tosini A. 2014  The Feynman problem and
  fermionic entanglement: Fermionic theory versus qubit theory. {\em
  International Journal of Modern Physics A} \textbf{29}, 1430025.

\bibitem{PhysRevLett.106.010501}
D'Ariano GM, Facchini S, Perinotti P. 2011  No Signaling, Entanglement
  Breaking, and Localizability in Bipartite Channels. {\em Phys. Rev. Lett.}
  \textbf{106}, 010501.

\bibitem{PhysRevA.77.062112}
Ziman M. 2008  Process positive-operator-valued measure: A mathematical
  framework for the description of process tomography experiments. {\em Phys.
  Rev. A} \textbf{77}, 062112.

\bibitem{Jencova:2012aa}
Jen{\v c}ov{\'a} A. 2012  Generalized channels: Channels for convex subsets of
  the state space. {\em Journal of Mathematical Physics} \textbf{53}, 012201.

\bibitem{selinger2008lecture}
Selinger P. 2008  Lecture notes on the lambda calculus. {\em arXiv preprint
  arXiv:0804.3434}.

\bibitem{selinger2004towards}
Selinger P. 2004  Towards a quantum programming language. {\em Mathematical
  Structures in Computer Science} \textbf{14}, 527--586.

\bibitem{selinger2009quantum}
Selinger P, Valiron B et~al. Quantum lambda calculus. {\em Semantic Techniques
  in Quantum Computation} pp. 135--172.

\bibitem{10.1007/11417170_26}
Selinger P, Valiron B. 2005  A Lambda Calculus for Quantum Computation with
  Classical Control. In Urzyczyn P, editor, {\em Typed Lambda Calculi and
  Applications} pp. 354--368 Berlin, Heidelberg. Springer Berlin Heidelberg.

\bibitem{hasuo2017semantics}
Hasuo I, Hoshino N. 2017  Semantics of higher-order quantum computation via
  geometry of interaction. {\em Annals of Pure and Applied Logic} \textbf{168},
  404--469.

\bibitem{malherbe2013categorical}
Malherbe O. 2013  Categorical models of computation: partially traced
  categories and presheaf models of quantum computation. {\em arXiv preprint
  arXiv:1301.5087}.

\bibitem{Pagani:2014:AQS:2578855.2535879}
Pagani M, Selinger P, Valiron B. 2014  Applying Quantitative Semantics to
  Higher-order Quantum Computing. {\em SIGPLAN Not.} \textbf{49}, 647--658.

\bibitem{SELINGER2007139}
Selinger P. 2007  Dagger Compact Closed Categories and Completely Positive
  Maps: (Extended Abstract). {\em Electronic Notes in Theoretical Computer
  Science} \textbf{170}, 139 -- 163.
Proceedings of the 3rd International Workshop on Quantum Programming Languages
  (QPL 2005).

\end{thebibliography}













\end{document}